\documentclass[twocolumn,tighten,usenatbib]{aastex631}
\usepackage{lmodern, fix-cm}
\usepackage{verbatim}
\usepackage{soul}
\usepackage{newtxtext,newtxmath}

\usepackage[T1]{fontenc}
\usepackage{ae,aecompl}
\usepackage{multirow}
\usepackage{graphicx}	
\usepackage{float}
\usepackage{amssymb}	
\usepackage[utf8]{inputenc}
\usepackage{mathtools}

\usepackage{color}      
\usepackage{placeins}
\usepackage{longtable, booktabs, threeparttablex,}

\usepackage[caption=false]{subfig}
\usepackage{blindtext}
\usepackage[normalem]{ulem}
\usepackage{subfiles}
\usepackage{hyperref}

\newcommand{\mi}[1]{\textsf{m12i}}
\newcommand{\mb}[1]{\textsf{m12b}}
\newcommand{\Msol}[1]{M\textsubscript{\(\odot\)}}

\newcommand{\upenn}{Department of Physics \& Astronomy, University of Pennsylvania, 209 S 33rd St, Philadelphia, PA 19104, USA}
\newcommand{\cca}{Center for Computational Astrophysics, Flatiron Institute, 162 5th Ave, New York, NY 10010, USA}
\newcommand{\columbia}{Department of Astronomy, Columbia University, 550 West 120th Street, New York, NY 10027, USA}
\newcommand{\ucdavis}{Department of Physics \& Astronomy, University of California, Davis, CA 95616, USA}

\shorttitle{stream--subhalo interactions}

\shortauthors{Arora et al.}

\begin{document}

\title{LMC-driven anisotropic boosts in stream--subhalo interactions}

\correspondingauthor{Arpit Arora}
\email{arora125@sas.upenn.edu}

\author[0000-0002-8354-7356]{Arpit Arora}
\affiliation{\upenn}

\author[0000-0001-7107-1744]{Nicol\'as Garavito-Camargo}
\affiliation{\cca}

\author[0000-0003-3939-3297]{Robyn E. Sanderson}
\affiliation{\upenn}
\affiliation{\cca}

\author[0000-0002-6993-0826]{Emily C. Cunningham}
\altaffiliation{NASA Hubble Fellow}
\affiliation{\columbia}
\affiliation{\cca}

\author[0000-0003-0603-8942]{Andrew Wetzel}
\affiliation{\ucdavis}

\author[0000-0001-5214-8822]{Nondh Panithanpaisal}
\affiliation{\upenn}
\affiliation{Carnegie Observatories, 813 Santa Barbara St, Pasadena, CA 91101, USA}
\affiliation{TAPIR, California Institute of Technology, Pasadena, CA 91125, USA}

\author[0000-0002-1176-0078]{Megan Barry}
\affiliation{\ucdavis}

\begin{abstract}
Dark matter subhalos are predicted to perturb stellar streams; stream morphologies and dynamics can, therefore, constrain the mass distribution of subhalos. Using FIRE-2 simulations of Milky Way-mass galaxies, we demonstrate that the presence of an LMC-analog significantly changes stream-subhalo encounter rates. The LMC-analog brings in many subhalos, increasing encounter rates for streams near the massive satellite by 10–40\%. Additionally, the LMC-analog displaces the host from its center-of-mass (inducing reflex motion), which causes a north-south asymmetry in the density and radial velocity distributions of subhalos. This asymmetry, {combined with the presence of LMC-analog subhalos}, causes encounter rates at the same distance to vary by 50–70\% across the sky, {particularly in regions opposite the LMC-analog}. Furthermore, the LMC-analog induces a density wake in the host’s DM halo, further boosting the encounter rates near the LMC-analog. We also explore how stream orbital properties affect encounter rates, finding up to a 50\% increase for streams moving retrograde to the LMC-analog's orbit in the opposite quadrant. {Finally, we report the encounter rates for Milky Way streams within the context of our simulations, both with and without the presence of an LMC-analog.} The dependence of encounter rates on stream location, orbit, and their position relative to the LMC has important implications for where to search for streams with spurs and gaps in the Milky Way.
\end{abstract}


\section{Introduction}\label{sec:intro}

A promising indirect method to constrain the nature of dark matter (DM) is by measuring the mass function of dark subhalos. At these small scales, different DM models (that have yet to be ruled out observationally) have different predictions for the number of expected subhalos as a function of mass. For example, in the cold dark matter (CDM) paradigm, the predicted number of dark subhalos {with a mass of about $10^7$ \Msol{}} for a galaxy like the Milky Way (MW) is approximately $10^3$ subhalos, while in warm dark matter (WDM) models the expected number depends on the mass of the WDM particles and ranges from $10 - 10^3$ dark subhalos \citep[e.g.,][]{kim2018missing}. In self-interacting DM models with velocity dependent cross-section and maximum transfer cross-section of about 3.5 cm$^2$g$^{-1}$, we expect around 100 subhalos \citep[e.g.,][]{vogelsberger2012subhaloes, robles2019milky, nadler2021effects}. Thus, measuring the abundance of subhalos at mass scales below $10^7$ \Msol{} would provide strong constraints between different DM models.
A promising way to detect subhalos in the MW is by observing their signatures after they interact with cold substructures such as stellar streams \citep{johnston2002lumpy, erkal2015properties, sanders2016dynamics, bovy2017linear, malhan2021probing}. The prime example is the GD-1 stream, a long (10 kpc, approximately 100$^{\circ}$) and thin (20 pc) stream around the MW \citep{grillmair2006detection}. The density of the stream is not smooth, but rather shows several gaps in density, one overdensity, and stars orbiting with the stream above the main stream track, known as ``the spur'' \citep{price2018off, malhan2019butterfly}. Detailed orbital modeling has constrained the mass of the perturber to be $10^6 - 10^8$ \Msol{}, possibly originating as a subhalo brought by the Sagittarius dwarf galaxy \citep{bonaca2019spur}. However, the concentration of this perturber is marginally consistent with predictions for CDM subhalos \citep{de2020closer, banik2021evidence, banik2021novel}. These results have motivated searches for further observational evidence of stream--subhalo interactions across the MW's halo.

Currently, there are about 100 known streams in the MW \citep{mateu2023galstreams} located at different distances and in different regions of the sky, providing a unique opportunity to detect the signatures of dark subhalos across the galaxy. Surveys such as Gaia \citep{gaia2016}, DESI \citep{desicollaboration2016desi}, H3 \citep{conroy2019H3}, the Vera Rubin Observatory \citep{ivezic2019lsst}, WEAVE \citep{dalton2012weave}, 4most \citep{de20194most}, and Subaru PFS \citep{takada2014extragalactic}, among others, will further observe streams all the way to the edge of the galaxy and in external galaxies \citep{pearson2022hough, aganze2023prospects}. The data from all of these surveys will provide a multidimensional view (kinematics and chemistry) of the stellar halo that will allow the detection and characterization of the morphology of stellar streams and, hence, the detection low-mass, completely dark subhalos.

Assuming that more perturbed streams can be found, and their perturbers constrained with dynamical modeling, obtaining constraints on the properties of the DM particle from these few perturbers requires interpreting their consistency with the population of subhalos predicted for each DM model. The usual practice currently is to assume that the DM subhalo population is isotropic \citep[e.g.][]{carlberg2009star, yoon2011clumpy, bovy2016detecting, erkal2015forensics, sanders2016dynamics}. However, the Milky Way's distribution of satellite galaxies and subhalos is almost certainly \textit{anisotropic} \citep[e.g][]{nadler2020milky,pawlowski2020milky, li2021gaia, savino2022hubble}. One of the main causes of this anisotropic distribution is the anisotropic accretion through filaments \citep{libeskind2011preferred} and recent accretion of the Magellanic Clouds, here we will focus on the effect from these massive satellites, which are causing several disequilibria throughout the galaxy \citep[e.g., see ][for a recent review]{vasiliev23review}.  In short, the LMC brings its own subhalos distributed along the DM debris \citep[e.g.,][]{deason2015satellites, wetzel2015satellite,sales2016identifying}, displaces the host from its center-of-mass (inducing reflex motion), causing a north-south asymmetry in the density and radial velocity distributions \citep[e.g.,][]{garavito2019hunting, petersen2020reflex, erkal2019total} , and induce a DM density wake in the host’s DM halo \cite{garavito2019hunting} and stellar halo \citep{cunningham2020quantifying, conroy2021all}. As a result, streams located in different locations in the sky but at the same galactocentric distance would experience different interaction rates with DM subhalos.

Recently, \cite{barry2023dark} calculated the enhancements effect of LMC--mass satellites on several subhalo population metrics, including the number density and orbital flux. They found that the presence of an LMC--mass satellite enhances these population metrics by up to factors of 1.2--2. {They also computed the spherically averaged subhalo encounter rates as a function of distance from the host center. However, within the MW, the encounter rates are subject to azimuthal variations due to the host's response to the in-falling satellite and the subhalos brought in by the LMC itself.} In this paper, we quantify the effect of massive LMC--mass satellite on the subhalo--stream interaction in one of the systems which is most LMC--like in its orbit and mass ratios at first pericenter. By using the Latte suite of FIRE-2 simulations \citep{wetzel2016reconciling, wetzel2023public}, we predict which regions in the halo would have higher subhalo-stream interactions. {Our study provides a more detailed and context-specific analysis of the MW system, offering a complementary perspective to the broader picture painted by \citet{barry2023dark}.}

The structure of this paper is as follows: In Sec.~\ref{sec:method}, we provide a detailed explanation of our choice of zoomed-in cosmological FIRE-2 simulations. We identify the closest MW-LMC analog in FIRE-2 suite and establish the analytical models to compute stream--subhalo encounter rates. Additionally, we demonstrate how we inject both the real MW and synthetic streams and integrate their orbits to calculate encounter rates. In Sec.~\ref{sec:anly}, we investigate the impact of in-falling satellites and the MW's response on encounter rates. We analyze key parameters influencing these rates, including the DM wake, collective response, and reflex motion. In Sec.~\ref{sec:results}, we present the computed encounter rates for both synthetic streams and the injected real MW streams. Statistical analysis is used to highlight key differences observed in the encounter rates, which are dependent on stream properties and their local positions. Finally, in Sec.~\ref{sec:MW_cluster}, we discuss the implications of our results in the context of the MW. Our conclusions are presented in Sec.~\ref{sec:conc}.

\section{Stream-subhalo encounters in simulations}\label{sec:method}
In this section, we detail our approach to studying the stream--subhalo encounter rates in simulations of MW-mass galaxies. We begin by explaining our choice of the simulations and the methodology for identifying the LMC-analog. Next, we establish the Galactocentric and stream-centric coordinate systems necessary for the analysis. We then outline our subhalo selection criteria and track the LMC-associated subhalos, examining their spatial distribution and assessing their ``survival status'', i.e., whether they are bound to or disrupted by the LMC-analog or the MW.

Furthermore, we introduce our analytical model, outlining the assumptions used to compute the stream--subhalo encounter rates. This involves integrating representative orbits for a suite of both synthetic and real MW streams.  

\subsection{Simulations}
\label{ref:simulations}

We select two cosmological zoomed-in baryonic simulations of MW-mass galaxies from the \textit{Latte} suite \citep{wetzel2023public} of the Feedback In Realistic Environments (FIRE) project.\footnote{\url{http://fire.northwestern.edu/latte}} These simulations are run with the FIRE-2 physics model \citep{hopkins2018fire} using the GIZMO code \citep{hopkins2015code} which utilizes a TREE+PM solver for gravity and a Lagrangian meshless-finite-mass (MFM) solver for hydrodynamics with adaptive spatial resolution. The FIRE-2 model implements star formation and stellar feedback parameters from the STARBURST99 stellar evolution models (\citealt{leitherer1999starburst99}) in the $\Lambda \textrm{CDM}$ cosmology from Planck \citep{collaboration2015planck}. A detailed description of the FIRE-2 project can be found in \cite{hopkins2018fire} and about the \textit{Latte} suite of MW-mass systems specifically in \citet{wetzel2016reconciling}. \citet{horta2023observable} made predictions for observable for the LMC-analog accretion events in the \textit{Latte} MW analogs while \citet{panithanpaisal2021galaxy} studied mass distribution of massive stellar stream and their progenitors. \citet{Samuel2020} examined radial velocity distribution of satellites around isolated and paired MW--M31 analogs and showed that central disc tidally destroys satellites altering their radial profile. \cite{samuel2021planes} showed that spatially thin MW-like planes of satellites can exist in cosmological simulation and the presence of an LMC-analog increases the probability of such planes. \cite{barry2023dark} did a statistical study of spherically average subhalo number density and estimated a boost of 1.2--2 due to the LMC-analogs. They also estimated that a stream like GD-1 \citep{grillmair2006detection} can have roughly 5--6 encounters per Gyr.

We select two galaxies labeled \mi{} and \mb{} for our analysis. Both are similar to the MW in stellar and gas mass content as well as their DM mass and density profiles at the present day \citep{hopkins2018fire, garrison2018origin, sanderson2020synthetic}. Each DM halo has a total mass of about $1. \times 10^{12}$ \Msol{}, and each simulation uses an initial particle mass of $m_\mathrm{b} = 7100$ \Msol{} for stars and gas, and dark matter particle mass $m_\mathrm{DM} = 35000$ \Msol{}. \mb{} includes a massive satellite analogous to the LMC in the MW with it's first pericenteric passage about 5 Gyr before the present day, while \mi{} has a relatively quiescent history with no major mergers for the past 8 Gyr and is used as our ``control'' system.

We identify the LMC-analog in \mb{} based on the similarity of its orbit around first pericenter (Fig.~\ref{fig:LMC_traj_m12b}), and on the similar merger mass ratio of about 1:8 at pericenter (Table 1). The orbit has $d_\textrm{peri} = 37.9$~kpc and $v_\textrm{tot} \approx 350$~km s$^{-1}$, comparable to the values of $d_\textrm{peri} = 49$~kpc and $v_\textrm{tot}$=350 km s$^{-1}$ estimated for the LMC \citep{kallivayalil2006proper}. Following \cite{arora2022stability}, we define the \textit{total mass ratio} (TMR) to be the mass of the host galaxy divided by the mass of the satellite at $T_\textrm{peri}$ and the pericenter mass ratio (PMR) to be the mass of the MW enclosed within $d_\mathrm{peri}$ divided by the mass brought in by the satellite within $d_\mathrm{peri}$ of the center of the satellite. The LMC-analog has a TMR of 7.3, while the estimated TMR for the LMC is about $\sim 4-15$, assuming a MW mass of $1-1.5 \times 10^{12} \ M_\odot$ and the LMC mass of $1-2.5 \times 10^{11} \ M_\odot$ \citep[e.g.][]{penarrubia2015_LMC, vasiliev2021tango}. 

The main difference between our analog and the LMC, based on the models in \cite{garavito2019hunting}, is the plane of the satellite's orbit relative to the host disk and that it's first pericentric passage occurred about 5 Gyr before the present day. In our simulation, the LMC-analog's orbit is inclined at $107^\circ$ oriented with respect to the disk plane. However, we do not expect this difference to significantly impact our conclusions. This is because we are interested in computing subhalo stream interactions and determining their rates in different halo quadrants relative to the merger trajectory. Moreover, since the orientation of the LMC orbit and the orbit is not very well-constrained due to uncertainties in measured positions and proper motion and in the MW mass models \citep{mcmillan2016mass, pietrzynski2019distance}, our simulation with a slightly differently oriented LMC-analog provides a plausible scenario. The torque from the disk could affect the orientation and shape of the streams, but the effect is expected to be minor. In fact, a number of studies have investigated the effect of the disk's torque on the shape of tidal streams. For example, \cite{law2005two} found that the disk torque can significantly affect the orientation and shape of streams for nearly coplanar orbits, while for more inclined orbits the effect is much weaker. On the other hand, \cite{kazantzidis2008cold} argued that the disk torque is generally not important for the formation of streams, since the subhalos motion in the host halo is dominated by the overall potential rather than the disk torque. \citet{santistevan2023modeling} showed that orbits of satellites in idealized host potentials are completely insensitive to host disk orientation. Furthermore, \cite{garrison2017not} showed that tidal disruption of subhalos remains largely unaffected by the geometry of the host disc. Therefore, while the effect of the disk torque on the shape of streams is still debated and likely depends on various factors such as the orbital parameters and the properties of the host and subhalo, it is generally believed to be small for orbits that are not nearly coplanar with the host disk. 

The first pericenter passage of the LMC-analog in \mb{} occurs at a lookback time of about 5 Gyr (Table.~\ref{tab:merger}) so we define the time $T$ relative to it such that $T_\textrm{peri} = 0$ Gyr. The most massive perturber in \mi{} reaches the pericenter around the same time and is nearly an order of magnitude less massive relative to its host at the pericenter than the LMC-analog we identify in \mb{}. The age of our host galaxies at pericenter is thus less than the current age of the MW; however, given that the majority of star formation has slowed down by about 7 Gyr in both \mb{} and \mi{} \citep{hopkins2018fire} and the total mass of each halo is close to the MW at the time of the merger, we expect our simulations to be analogous to the real MW even though they are slightly younger than the MW at satellite pericenter. Table.~\ref{tab:merger} summarizes the characteristics of the LMC-analog in \mb{} and the equivalent most massive satellite in \mi{} at $T_\textrm{peri} = 0$ Gyr.   

\citet{barry2023dark} showed that the presence of an LMC-analog can boost the subhalo number density by a factor of 1.2--2. They identified four LMC-analogs in simulations \mb{}, m12f, m12w, and m12c from the Latte suite of FIRE-2 simulations, approximating properties when the analogs approach within 50 kpc of the center and not their actual pericentric distances. This method provides greater statistical accuracy when evaluating the contribution of the LMC-analogs to the subhalo population, but not necessarily on subhalo phase-space distribution. Our focus here encompasses a global perspective that also considers the effects of the MW response to the in-falling satellites, which is dependent on the satellite's orbit, actual pericenter distance, mass, and velocity. The LMC-analog orbit in m12f is similar to the expected LMC orbit, but the in-falling satellite is only half as massive, with a TMR of about 16 \citep{arora2022stability}. Consequently, the host's response to the LMC-analog in m12f is significantly weaker.

Moving on to m12w and m12c, both simulations feature satellites similar in mass to the LMC. However, m12w stands out due to its highly eccentric, fast-moving orbit, resulting in a very radial merger with its first pericenter at 8 kpc and eventual complete tidal disruption of the satellite. It's almost 6.5 Gyr before the present day (1.5 Gyr earlier in comparison to \mb{}). This makes it highly unreliable for predictions. In contrast, the LMC-analog in m12c is closest to the present day but follows an orbit completely within the disc plane, with the actual first pericenter occurring at 18 kpc. This particular orbit is not expected to induce a north-south asymmetry. Selecting an arbitrary pericenter at 50 kpc would not accurately represent halo deformations. For the sake of completeness, we have included our encounter rates calculation for m12c in appendix~\ref{app:m12c_enc}. 

\begin{table*}
\caption{Properties of the most massive mergers in \mb{} (LMC-analog) and \mi{}.}
\begin{center}
\begin{tabular}{c|cccccccc}
\multicolumn{1}{l}{\textbf{Simulation}} & \multicolumn{1}{l}{\textbf{\begin{tabular}[c]{@{}l@{}} $\mathbf{T_{peri}}$ \\ {[}Gyr{]}\end{tabular}}} & \multicolumn{1}{l}{\textbf{redshift}} & \multicolumn{1}{l}{\textbf{\begin{tabular}[c]{@{}l@{}} $\mathbf{M_{\star,host}}$ \\ {[}$\times 10^{10}$ \Msol{}{]}\end{tabular}}} & \multicolumn{1}{l}{\textbf{\begin{tabular}[c]{@{}l@{}} $\mathbf{d_{\mathrm{peri}}}$ \\ {[}kpc{]}\end{tabular}}} & \multicolumn{1}{l}{\textbf{TMR}} & \multicolumn{1}{l}{\textbf{PMR}} & \multicolumn{1}{l}{\textbf{\begin{tabular}[c]{@{}l@{}}$\mathbf{v_{\mathrm{rad}}}$ \\ {[}km s$^{-1}${]}\end{tabular}}} & \multicolumn{1}{l}{\textbf{\begin{tabular}[c]{@{}l@{}}$\mathbf{v_{\mathrm{tan}}}$ \\ {[}km s$^{-1}${]}\end{tabular}}} \\ \hline \hline
\mi{} & 8.05 & 0.60 & 3.8 &29.53 & 45.5 & 14.05  & -13.4 & 290.2 \\ \hline
\mb{} & 8.81 & 0.49 & 6.3 &37.9 & 8 & 3  & -9.7 & 349.2 \\ \hline
\end{tabular}
\end{center}
\tablecomments{$T_{\mathrm{peri}}$: Time of closest approach (``pericenter'') between the main galaxy and the satellite ($T = 0$ Gyr). All properties are evaluated at $T_{\mathrm{peri}}$.
 $M_{\star,host}$: Stellar mass of the halo. $d_{\mathrm{peri}}$: Pericenter distance between the satellite and the main galaxy. TMR: Total Mass Ratio of the MW and the satellite, $M_{\mathrm{main}}/M_{\mathrm{sat}}$, and PMR: Pericenter Mass Ratio, $M_{\mathrm{main}}(<d_{\mathrm{peri}})/M_{\mathrm{sat}}(<d_{\mathrm{peri} \mathrm{\, from \, sat}})$. $v_{\mathrm{rad}}$, $v_{\mathrm{tan}}$: radial and tangential velocities of the satellite with respect to the MW. $v_\mathrm{rad}$ is non-zero due to finite time resolution between snapshots around the pericentric passage.}
\label{tab:merger}
\end{table*}

\subsection{Coordinates and frames for analysis} \label{Sec:coord_Vdist}

The MW--LMC system we consider involves two galaxies orbiting around their common center of mass (COM), which is moving with relatively constant velocity through space in a cosmological simulation box. The frame centered on and moving with the system COM is thus an inertial frame (modulo interactions with the next most massive satellite galaxy of the MW). We will refer to this as the {system frame}, {marked by primed quantities}, related to the one in which the simulation is run (the ``simulation box frame''). 

From our perspective in the MW's disk, however, the Sun orbits the center of the MW (the local minimum in the potential and maximum in density), and we commonly use a coordinate system centered on and moving with this location, referred to as the \textbf{Galactocentric frame}. This frame is not inertial, since the MW and the LMC are orbiting their common COM, but instead is related to the system frame by: 
\begin{eqnarray}
    \vec{x} = \vec{x}' - \vec{x}'_{\mathrm{MW}}(t) \\
    \vec{v} = \vec{v}' - \vec{v}'_{\mathrm{MW}}(t)
\end{eqnarray}
where the unprimed quantities are in the Galactocentric frame, and $\vec{x}'_\mathrm{MW}$ and $\vec{v}'_\mathrm{MW}$ denote the position and velocity of the MW's center in the system frame, both of which are functions of time (as explicitly highlighted in this equation). Since $d\vec{v}_\mathrm{MW}/dt \neq 0$, the Galactocentric frame is \emph{not} inertial. The displacement of the Galactocentric frame in position and velocity is defined so that at the time of pericenter $t_{\mathrm{peri}}$ (the present configuration in the actual MW-LMC system):
\begin{eqnarray}
    \vec{x}(t_{\mathrm{peri}}) = \vec{x}' -\vec{x}'_\mathrm{ref} \\
    \vec{v}(t_{\mathrm{peri}}) = \vec{v}' - \vec{v}'_{\mathrm{ref}} ,
\end{eqnarray}
where $\vec{x}'_\mathrm{ref}$ is the distance between the MW center and the system COM, and $\vec{v}'_{\mathrm{ref}}$ is the relative motion of the MW's center induced by the LMC also known as the ```reflex motion,'' which for the real MW--LMC system is approximately 30--40 km s$^{-1}$ at the location of the Sun \citep{erkal2019total, petersen2020reflex, vasiliev2021tango, vasiliev23review}. {The LMC also accelerates the subhalos orbiting in the MW globally, leading to a net bulk motion of these subhalos directed towards the MW-LMC center. The acceleration experienced by an individual subhalo varies depending on its location in the galaxy with respect to the LMC. The LMC would roughly equally accelerate the subhalos around a stream and the stream. However, if subhalos are anisotropically or asymmetrically distributed around the stream, small variations in the accelerations experienced by the subhalos due to the LMC will lead to a perturbatively induced non-zero bulk motion of the subhalos around the stream. In contrast, if the subhalos are isotropically distributed, the net bulk motion around a stream will be zero.}

{Given that the simulations are run in arbitrary simulation box frame, the traditional approach is to establish a \textbf{``principal axes''} by aligning the galactic disk to the XY plane at the time of the analog's first pericentric passage. In this frame, the LMC-analog in \mb{} is located at $\vec{x}_\textrm{analog} = (-23.3, -26.6,  13.7)$~kpc at $T_\textrm{peri}$. However we establish an additional ``rotated axes'', that aligns the position unit vector of the LMC-analog in \mb{} at its first pericentric passage with the position unit vector of the \emph{real} LMC in the MW at its first pericentric passage, located at $\vec{x}_\textrm{LMC} = (2.3, -20.2, -41.1)$~kpc, based on the orbits presented in \citet{garavito2019hunting}. Details on how to compute this rotation can be found in Appendix.\ref{app:lmc_rot}.}   

Fig.~\ref{fig:LMC_traj_m12b} plots the trajectory of the LMC-analog (red/green) in \mb{} in the {rotated axes} Galactocentric coordinates, along with the total distance from the MW's central location (on the bottom right), compared to an MW-LMC constrained simulation (blue) from \cite{garavito2019hunting}. The green sections of the trajectories indicate the periods over which we compute encounter rates. {The overall orbit and the present-day location of the analog  match reasonably well that of the LMC (see Fig.~\ref{fig:reflex_mot}). Unless otherwise noted, all analyses and results in this paper are presented in the rotated axes frame.}


\begin{figure}
    \includegraphics[width=\linewidth]{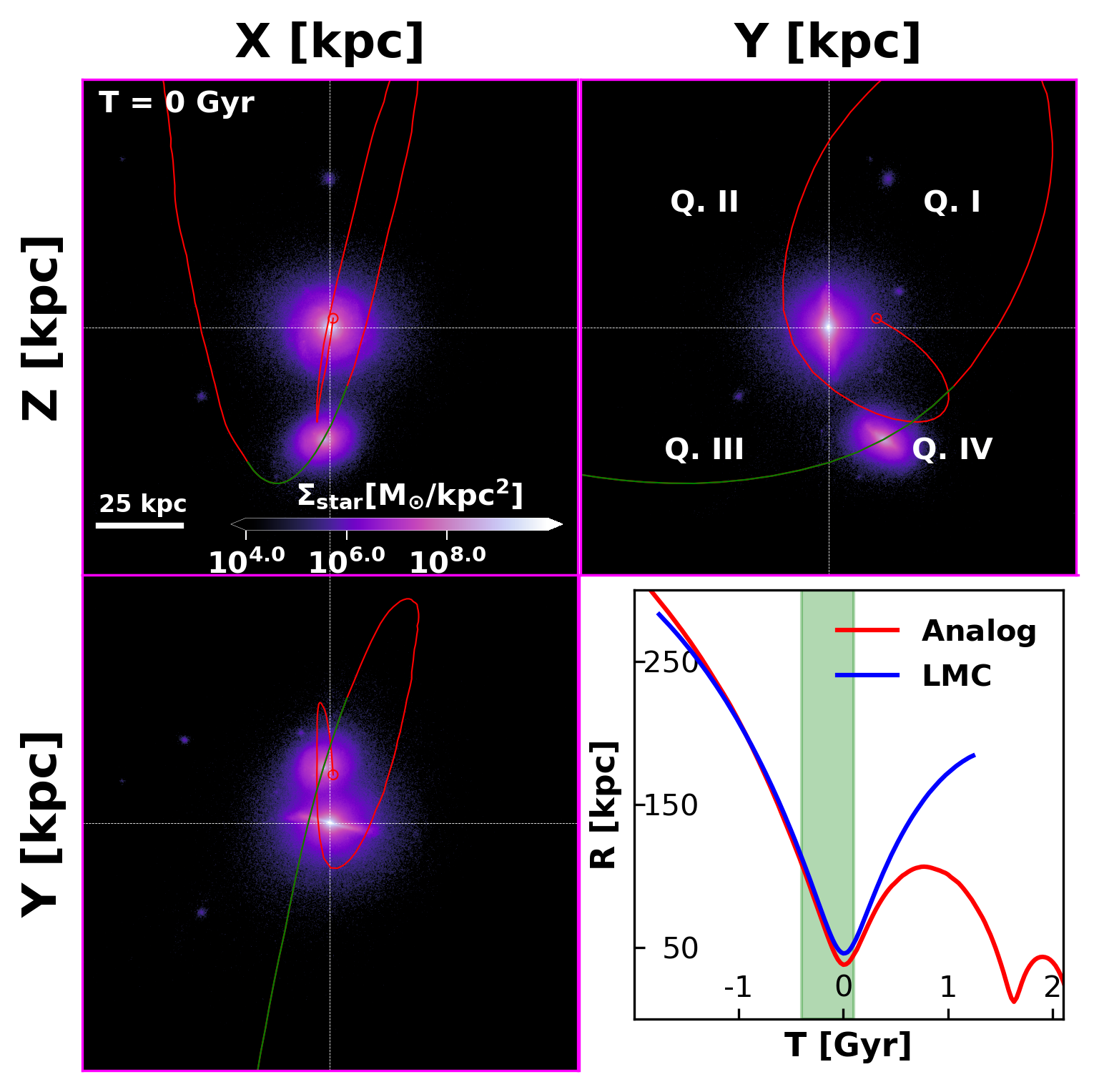}
        \caption{Trajectory of the LMC-analog (red/green) in the {rotated axes} Galactocentric coordinates from FIRE-2 simulation of a MW-mass galaxy labeled \mb{} with stellar surface density in the background at $T = 0$~Gyr (1st pericenter). Quadrant labels based on Table.~\ref{tab:quad_sky} are marked in the YZ plane. Bottom right: the distance from the galactic center as a function of time for the FIRE analog (red/green), compared to the MW-LMC simulation 3 from \citep{garavito2019hunting} (blue). {The green sections of the trajectories indicate the periods over which we compute encounter rates.}}
    \label{fig:LMC_traj_m12b}
\end{figure}

We will compute the subhalos-stream encounter rates for streams distributed in four quadrants in the sky. We define each quadrant in the {Galactocentric-rotated axes frame in} Table.~\ref{tab:quad_sky} for reference, also marked in the Fig.~\ref{fig:LMC_traj_m12b} YZ panel.

\begin{table}
\caption{Quadrant cutoffs in Sky coordinates.}
\begin{center}
\begin{tabular}{cccc}
\textbf{Quadrant} & \textbf{No.} & $\pmb{\phi} \ (^\circ)$ & $\pmb{\theta} \ \textrm(^\circ)$ \\ \hline \hline
North-East        & I            & -180 -- 0                  & 0 -- 90                             \\
North-West        & II           & 0 -- 180                    & 0 -- 90                             \\
South-West        & III          & 0 -- 180                     & -90 -- 0                             \\
South-East        & IV           & -180 -- 0                  & -90 -- 0         \\
\hline 
\end{tabular}
\end{center}
\tablecomments{where $\phi$ and $\theta$ correspond with azimuth and altitude respectively in mollweide projection. Refer to Fig.~\ref{fig:reflex_mot} b.) for a visual representation.}
\label{tab:quad_sky}
\end{table}

Finally, to compute quantities relevant to stream--subhalo encounters the relevant frame is the one moving with the stream, here referred to as the \textbf{stream-centric frame}. Since most of the currently known streams are closer to the MW center than the LMC, we will relate the stream-centric frame, denoted by $\tilde{\vec{x}}$ and $\tilde{\vec{v}}$, to the Galactocentric frame $\vec{x},\vec{v}$ (for streams orbiting beyond the LMC one would substitute the system frame $\vec{x}'$):
\begin{eqnarray}
    \tilde{\vec{x}} = \vec{x} - \vec{x}_{\mathrm{str}}(t) \\
    \tilde{\vec{v}} = \vec{v} - \vec{v}_{\mathrm{str}}(t), 
\end{eqnarray}
where $\vec{x}_{\mathrm{str}}(t), \vec{v}_{\mathrm{str}}(t)$ are the position and velocity of a representative particle in the stream in the Galactocentric coordinates (in other words, the stream's representative orbit). {The relevant quantity for determining the rate of subhalo interactions is the distribution of the subhalos velocities perpendicular to the stream that is the cylindrical radial velocities of subhalos in this stream-centric frame $\tilde{v}_R$.}

\subsection{Integrating stellar stream orbits} \label{Sec:injected_st}

In order to compute the encounter rates we need to inject stellar stream in the simulations and integrate their orbits, here we describe the methodology we employ.

\subsubsection{Stellar streams injection scheme}

Each stream is represented as a test particle that is integrated using the time-dependent potential of the system. We generate two sets of stream catalogs. One resembling the MW streams and one a synthetic catalog of 5000 streams.  The properties and ICs used to generate these two catalogs are described below: 

\begin{itemize}
    \item \textbf{MW streams}: 97 streams orbiting the MW with phase-space properties resembling the observed streams of the MW.  We use {\fontfamily{qcr}\selectfont galstreams}\citep{mateu2023galstreams} to compute the median phase-space values of each stream. {We then inject the streams positions and velocity to our rotated axes frame aligning the orientation of the streams and the LMC-analog in our model with that of the real LMC.}. 

    \item \textbf{Synthetic streams}: We generate 5000 synthetic stream test particles uniformly spread across the sky in a mollweide projection and different orbital properties. We simulate these streams to be at their pericenter with distances between 20-80 kpc from the galactic center at $T = 0$ Gyr, and varying azimuthal velocities and inclination velocities such that the total tangential velocities at $T = 0$ Gyr are between 250-400 km s$^{-1}$ (set by the real MW streams). These initial conditions produce a uniform distribution of streams' orbital eccentricities and pericentric distance in all quadrants. 
\end{itemize}

These ICs are used to start the orbital integration for each stream at $T = 0$ Gyr. We then integrate the orbits for a total of 0.5 Gyr corresponding to the shaded green band in Fig.~\ref{fig:LMC_traj_m12b} around the pericenter of the LMC-like satellite. {We integrate backward for -0.4 Gyr starting at T=0 Gyr and then forward by 0.1 Gyr starting from T=0 Gyr.} 0.5 Gyr is sufficient to induce noticeable morphological changes in a stream's structure, such as the emergence of open gaps and the occurrence of kinks \citep{erkal2015forensics}.
In the following section, we describe the time-dependent model that we use in our orbital integration. 

\subsubsection{Orbit integration of streams}

To integrate the orbits of the test particles (each representing a stream), we use time-evolving potential models from \cite{arora2022stability} following the scheme described in \citep{arora2024efficient}. In short, we fitted the potentials with basis function expansions (BFE) on the host halo at every snapshot in the simulations within the 20 snapshots that span the 0.5 Gyr around the  $T = 0$. These models can adequately describe deformations in the halo caused by the LMC-analogs \citep[e.g][]{petersen2020reflex, garavito2021quantifying, petersen2022exp, arora2022stability}. BFE have also been shown to successfully reproduce orbits, even in the presence of massive satellites, for short periods of time (less than 1 Gyr) in both idealized simulations \citep{lilleengen2023effect, vasiliev2023dear}, and cosmological simulations \citep{lowing2011halo, sanders2020models, arora2024efficient}. 

Since we represent each stream by the orbit of a single test particle, we do not simulate the perturbations along the leading and trailing arms of the stream (as is observed, for example, in the real-life Orphan stream \citep{lilleengen2023effect}).

The integration is performed in the principal axes frame on which the BFE models are fit. Nonetheless, we report our results in the rotated axes frame. Our orbit integration of subhalos and streams is not entirely self-consistent, as the subhalos respond to the LMC-analog in a live simulation, while the streams are integrated within a smooth, time-evolving potential model. However, one can simply integrate the subhalo orbits within the time-dependent potential using a prescription to account for tidally destroyed subhalos and count flybys. This approach is certainly feasible, but it also relies on the assumption that subhalo orbits can be properly reproduced, without taking into account the effects of dynamical friction.

We use the integrated orbits to calculate the apocenter and pericenter distances, and categorize the streams as prograde or retrograde relative to the orbit of the LMC-analog in \mb{}. Only about 13 of the known MW streams have pericenter distances greater than 20~kpc from the galactic center in the \mb{} potential.
           
\subsection{Subhalo selection}

We use the {\fontfamily{qcr}\selectfont ROCKSTAR} halo finder \citep{behroozi2012rockstar} to identify DM subhalos. Merger trees were constructed to link the catalogs at each snapshot using the {\fontfamily{qcr}\selectfont consistent-trees} code \citep{behroozi2012consistent}, as detailed in \cite{samuel2021planes} and \cite{panithanpaisal2021galaxy}. Using these halo catalogs we identify subhalos that {are within the MW and} brought in by the LMC-analog. {We impose a mass range of $10^6$ \Msol{} $\leq \mathrm{M_{sub}} \leq 10^9$ \Msol{} on all the subhalos based on our resolution.} We select the subhalos bound to and within the virial radius of the LMC-analog 2~Gyr before the first pericenter; when the $\mathrm{d_{LMC}} + \mathrm{R_{vir,LMC}}$ is greater than the $\mathrm{R_{vir,MW}}$, and track their evolution through time until disruption using {\fontfamily{qcr}\selectfont consistent-trees}. 

\begin{figure*}
    \includegraphics[width=\linewidth]{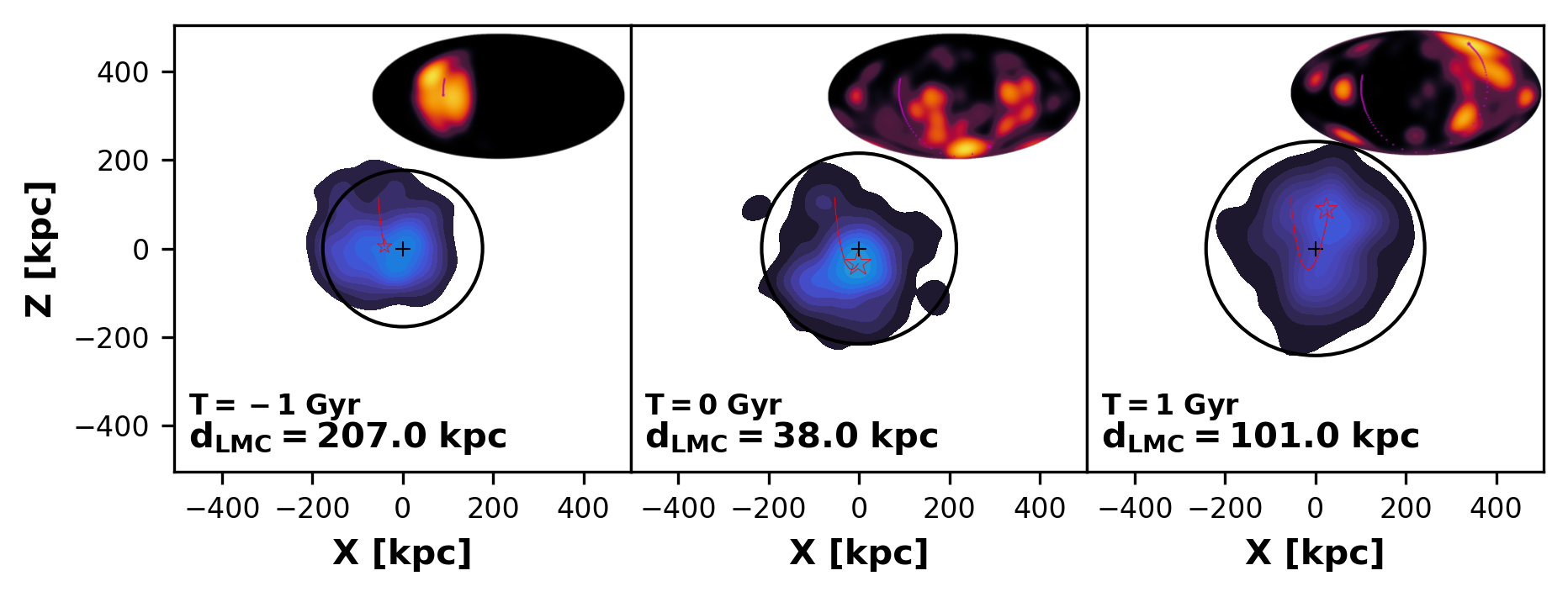}
    \caption{Density contours of the surviving subhalos brought in by the LMC-analog at three time steps: -1 Gyr (left), 0 Gyr (middle), and 1 Gyr (right) along with the analog's trajectory (red) in XZ plane for \mb{}. The black circle marks the virial radius of the MW, and the top right inset shows the LMC-analog's subhalo density in mollweide projection at each time step. As the analog falls in the MW, the LMC-analog subhalos are tidally stripped away from it and settle into bound orbits around the MW. This is obvious at 0 and 1 Gyr, as the density contours are more widespread in both cases. At 0 Gyr, Majority of the subhalos lag behind the in-falling LMC-analog, while some completely settled into newer orbits producing a dipole term in density. At 1 Gyr, almost all of the subhalos are dispersed away from the LMC-analog, as the central density of subhalos is away from the LMC-analog. Also note, there is slingshot action as a few subhalos end up unbound and outside the virial radius of the MW after the analog's flyby (top right corner for $T = 1$ Gyr).}
    \label{fig:dens_cont_m12b}
\end{figure*}

The LMC-analog brings a total of 1029 subhalos. Fig.~\ref{fig:dens_cont_m12b} plots the location of the LMC-analog subhalos as density contours in the XZ plane for \mb{} at -1~Gyr (left), 0~Gyr (middle), and 1~Gyr (right) with the LMC-analog trajectory and location (shown in red). Insets on top right shows the density of these subhalos in mollweide projection. The LMC-analog starts in the south and falls toward the galactic center bringing it's own population of subhalos that are widely distributed in different regions and settle into newer bound orbits around the MW as the merger proceeds. Note the trailing subhalos behind the satellite at pericenter in both the Cartesian plane and mollweide projection. Similar LMC subhalo distributions with leading and trailing arms are noted in \citet{sales2016identifying}.     

We track whether these subhalos are bound to the LMC-analog or the MW using a total negative energy criteria (i.e $E_\textrm{sub,tot} \leq 0$ in the LMC-analog or the MW reference frame). We also note whether the subhalos merge into the LMC-analog or the MW. A subhalo is considered ``merged'' when {\fontfamily{qcr}\selectfont consistent-trees} can no longer track it's center and it's descendant subhalo links to either the MW or the LMC-analog. {\fontfamily{qcr}\selectfont consistent-trees} can successfully track subhalos with masses greater than 10$^6$ \Msol{} \citep{samuel2021planes}. 
\begin{figure}
    \includegraphics[width=\linewidth]{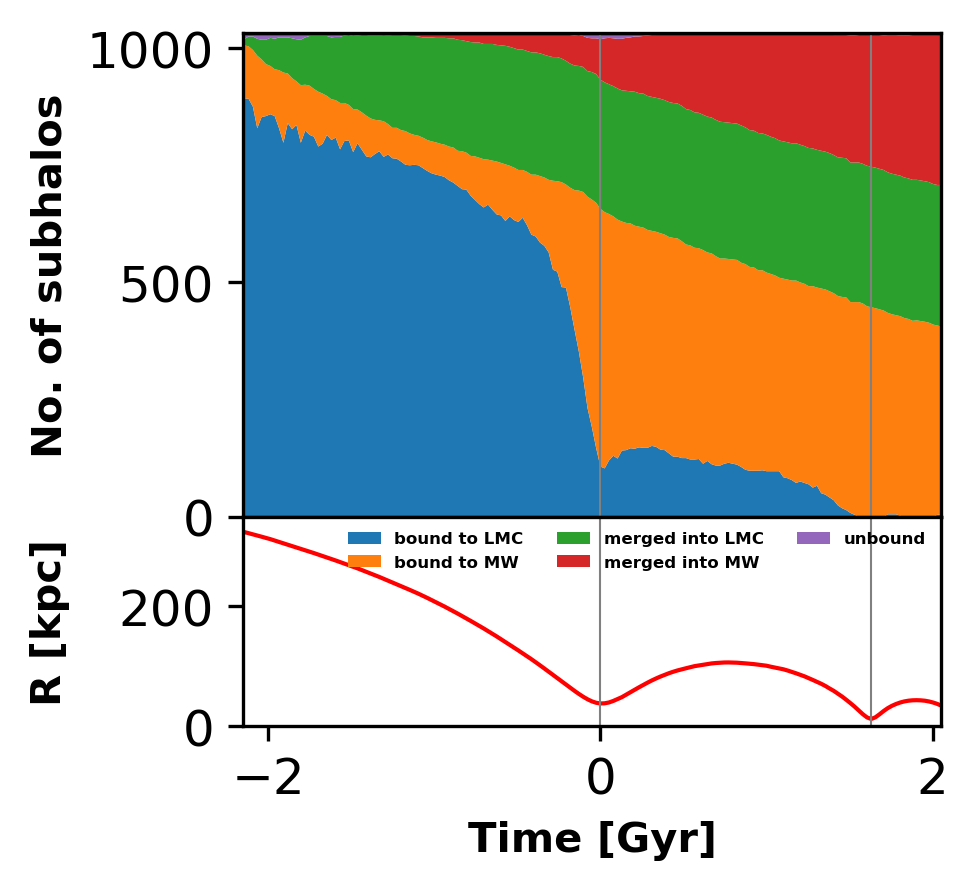}
    \caption{The bound properties of 1029 LMC-analog subhalos with mass range $10^6$ \Msol{} $\leq \mathrm{M_{sub}} \leq 10^9$ \Msol{} at different time steps to illustrate whether they are still bound to the LMC-analog (blue), bound to the MW (orange) or merged into the MW (red) or the LMC-analog (green). The number of subhalos bound to the analog drastically drops after the first pericenteric passage where the tidal forces from the the MW are stronger and hence they become bound to the MW. By the second pericenter passage, most of the subhalos are either bound to the MW or have been completely tidally stripped. Only about ~50\% subhalos survive after the merger is complete. The number of subhalos brought in by the LMC-analog that merged into it (green) increases as the analog gets closer to the MW but remains constant after the first pericenter. While more subhalos merge into the MW (red) as the merger proceeds. Note the few unbound subhalos at 0 Gyr are due to the instability of potential model to accurately describe the MW--LMC system.}
    \label{fig:subhal_bound}
\end{figure}

Fig.~\ref{fig:subhal_bound} illustrates the number of LMC-analog subhalos bound to the LMC-analog (blue) or the MW (orange), merged into the LMC-analog (green) or the MW (red) as a function of time starting from 2~Gyr before the first pericenter passage. By the first pericenter (0~Gyr), 80\% of the subhalos are tidally stripped from the satellite and bound to the MW. By 1 Gyr, only half of the subhalos survive the merger while the rest are destroyed and form dark streams. In the scope of our simulations and our inability to resolve these dark streams, we leave them out of our calculations for future research. These dark streams are less likely leave observable signatures after their interactions with stellar streams because of their low surface density. 

\subsection{Stream--subhalo encounter rates} \label{sec:anly_setup}

Counting the number of stream--subhalo encounters is not feasible due to the limited temporal resolution in the simulations. Our time range consists of approximately 20 snapshots, each spaced 25 Myr apart.  Therefore, we calculate the encounter rate at each snapshot from our simulation between the subhalos and a stellar stream using the counting scheme described by \citet{erkal2016number}, which is similar to the one conducted by \citet{yoon2011clumpy}. The encounter rates are computed as the number of subhalos entering a cylinder of radius $b_\textrm{max}$ (impact parameter) around a straight line stream of length $\ell_s$. The encounter rates depend on the local number density of subhalos $n_\textrm{sub}$ around the stream and the cylindrical radial velocity $\tilde{v}_R$ distribution of subhalos with respect to the stream. 

The number of subhalos passing through a cylinder of length $\ell_s$ and radius $b_\textrm{max}$ at time $t_\textrm{snap}$, within a time interval $dt$, is then given by:

\begin{equation} \label{eq:dNenc_raw}
    dN_\textrm{enc} (t_\textrm{snap}) = (2\pi b_\textrm{max} \ell_s) \cdot (|\tilde{v}_R| dt) \cdot n_\textrm{sub} \cdot P(\tilde{v}_R) d\tilde{v}_R
\end{equation}

Here, $P(\tilde{v}_R)$ represents the probability distribution function of $\tilde{v}_R$. {In contrast to \citet{erkal2016number}, where they considered a straight line stream of length $\ell_s$ and computed encounter rates by rates by modeling a cylinder of length $\ell_s$ and radius $b_\textrm{max}$ around the stream, our approach simplifies the stream to a single particle representation. This approximation does not accurately describe a cylinder around the stream. Instead, we calculate $\tilde{v}_R$ as the relative velocity component perpendicular to the tangent plane of the stream at $\hat{x}_\textrm{str}$ as  $\tilde{v}_R = ( \vec{v}_{\mathrm{sub}} - \vec{v}_{\mathrm{str}}) \cdot \hat{x}_\textrm{str}$ with direction towards the plane, both from the top and bottom of the plane represented with negative values. This approach overlooks the orientation of the stream on the tangent plane itself. Subhalos moving towards the plane with a zero component perpendicular to the plane can still perturb the stream, an effect not accounted for in our single particle model. Therefore, we integrate over different orientations of the plane assuming $P(\tilde{v}_R)$ remains symmetric under small rotations, imitating the change in direction of the normal to the cylinder. The produces an extra factor of $2\pi$. In this calculation, we consider only subhalos entering through the sides of the ``cylinder'', so we integrate over all negative $\tilde{v_R}$}. We can then re-write the encounter rates as:

\begin{eqnarray} 
    \frac{dN_\textrm{enc}}{dt} (t_\textrm{snap}) &=& (2\pi b_\textrm{max} \ell_s) \cdot n_\textrm{sub} \cdot \mathcal{I}(\tilde{v}_R) \label{eq:dNenc_dt} \\    
    \mathcal{I}(\tilde{v}_R) &\equiv& \int_{-\infty}^{0} |\tilde{v}_R| \cdot P(\tilde{v}_R) d\tilde{v}_R \label{eq:I_vR}
\end{eqnarray}

{Where $\mathcal{I}(\tilde{v}_R)$ is the first  moment of the probability distribution of subhalos entering the cylinder ( $\tilde{v}_{R} \leq 0$). $P(\tilde{v}_R)$ is often approximated by a Gaussian distribution of $\tilde{v}_R$ assuming a non-zero mean ($\tilde\mu$) and dispersion ($\tilde\sigma$) \citep{yoon2011clumpy, erkal2016number}. In such a scenario, $\mathcal{I}(\tilde{v}_R)$ simplifies to:} 

\begin{equation} \label{eq:num_gauss}
\begin{aligned}
    \mathcal{I}_\mathrm{gauss}(\tilde{v}_R) = \frac{\tilde\sigma}{\sqrt{2\pi}} \cdot f(\tilde\gamma)\\
    \textrm{where,} \ \  f(\tilde\gamma) & \equiv  \textrm{e}^{-\tilde\gamma^2} + \sqrt{\pi}\tilde\gamma (\textrm{erf}(\tilde\gamma)-1),
\end{aligned}  
\end{equation}

where $\tilde\gamma \equiv \frac{\tilde\mu}{\sqrt{2}\tilde\sigma}$. $f(\tilde\gamma)$ is a unitless scaling factor that depends on the mean and dispersion of  $\tilde{v_R}$, hereafter the \textbf{\textit{anisotropic boost factor}}. This factor equals unity when $\tilde\mu = 0$ {(see Fig.~\ref{fig:gamma_scal_fact} in Appendix~\ref{app:gaussianity} for more details)} for the traditional encounter rate formulation such as in \citet{yoon2011clumpy, erkal2016number, barry2023dark}. {In Appendix~\ref{app:gaussianity}, we demonstrate that this approximation is valid to the within 10\% error compared to the numerical integration.} 
 
The density of subhalos, $n_\textrm{sub}$, is computed as a function of time by counting the number of subhalos within a 10 kpc width spherical shell centered around the stream. {This shell is further confined to a sky slice that spans $\pm \pi/2$ in colongitude and $\pm \pi/4$ in colatitude around the stream's position. The stream is always positioned at the center of this slice at each time step. We then divide the counts by the volume of this sliced shell, which is one-fourth of the total volume of the spherical shell. Our decision not to use a local volume around the stream was made to mitigate numerical noise owing to low subhalo counts. Similarly, we evaluate $P(\tilde{v}_R)$ from subhalos within the same shell, and in the stream's quadrant for each time step of the stream. We numerically integrate eq.~\ref{eq:I_vR} using $P(\tilde{v}_R)$ to compute the encounter rates for each time snapshot using eq.~\ref{eq:dNenc_dt}.}  

Finally, we sum the number of encounters per Gyr for a stream using the equation:

\begin{equation}\label{eq:D_N/D_T}
    \frac{N_{\mathrm{enc}}}{\mathrm{Gyr}}=\frac{\sum_{t_\textrm{snap}} \frac{d N_{\mathrm{enc}}}{d t} (t_\textrm{snap}) \Delta t_\textrm{snap}}{\sum_{t_\textrm{snap}} \Delta t_\textrm{snap}},
\end{equation}

Where, $\Delta t_\textrm{snap}$ represents the time interval between snapshots, and the sums are taken over all time steps $t_\textrm{snap}$ and $\dfrac{d N_{\mathrm{enc}}}{dt}$ is computed using eq.~\ref{eq:dNenc_dt}  . 

{In this paper we don't explore the impacts caused by any single encounters in our analysis which are more sensitive to the individual stream and subhalo kinematics and require a case-by-case study. We rather focus on a global analysis of the encounter rates in this paper. Traditionally an impact-weighting based on the impulse approximation $\sim \frac{G M_\textrm{sub}}{v^2 b}$, where $b$ is the distance of the closest approach between a subhalo and a stream has been used in literature \citep{erkal2015forensics, erkal2016number}. However, implementing such weighting in our method would require integrating over the mass distribution of subhalos, which is trivial since the $P(M_\textrm{sub}) \propto 1/M_\textrm{sub}$, and a distribution of $b$. Finding $b$ requires computing pair-wise distance between each stream and all the subhalos within some set distance cut off around the stream during the integration time and the number of subhalos changes non-trivially. Alternatively, one can assume $b$ is a non-linear function of the velocity perpendicular ($\tilde{v}_R$) and parallel ($\tilde{v}_{||}$) to the stream. These assumptions lead to a coupled 2D integral with both the velocity distributions in the denominator and $\sqrt{\tilde{v}_{R}^2 + \tilde{v}_{||}^2}$ in the numerator, this integral diverges at 0 and is highly sensitive to minor fluctuations. However, in our case the impact-weighting for the subhalos brought in by the LMC-analog will be similar to the impact-weighting of the MW subhalos, this is demonstrated by showing that the velocity field is fairly independent of the  subhalo mass in Sec.~\ref{sec:anly}}.

\section{Effects of massive satellites on encounter rates}\label{sec:anly}
In this section, we illustrate how and why each of the physical quantities in the encounter rates (subhalos number density and radial velocity distribution) are affected during the infall of the LMC-analog. In Section~\ref{sec:phase-space} we start by exploring the temporal evolution of the phase-space distribution of the host and contributed subhalos. {In Section~\ref{sec:velocity_changes} we explore the temporal changes in the radial velocity distribution of the subhalos caused by the perturbations of the LMC-analog. Afterward in Section~\ref{sec:radial_dependence} we explore the radial and angular variations on the sky
in the integral of the probability distribution $\mathcal{I}({v}_\textrm{rad})$.} Lastly, we quantify the evolution of the number density of subhalos in Section~\ref{sec:subhalos_density}. 

\subsection{Phase-space distribution of the perturbed subhalo population}\label{sec:phase-space}

\begin{figure*}
    \includegraphics[width=\linewidth]{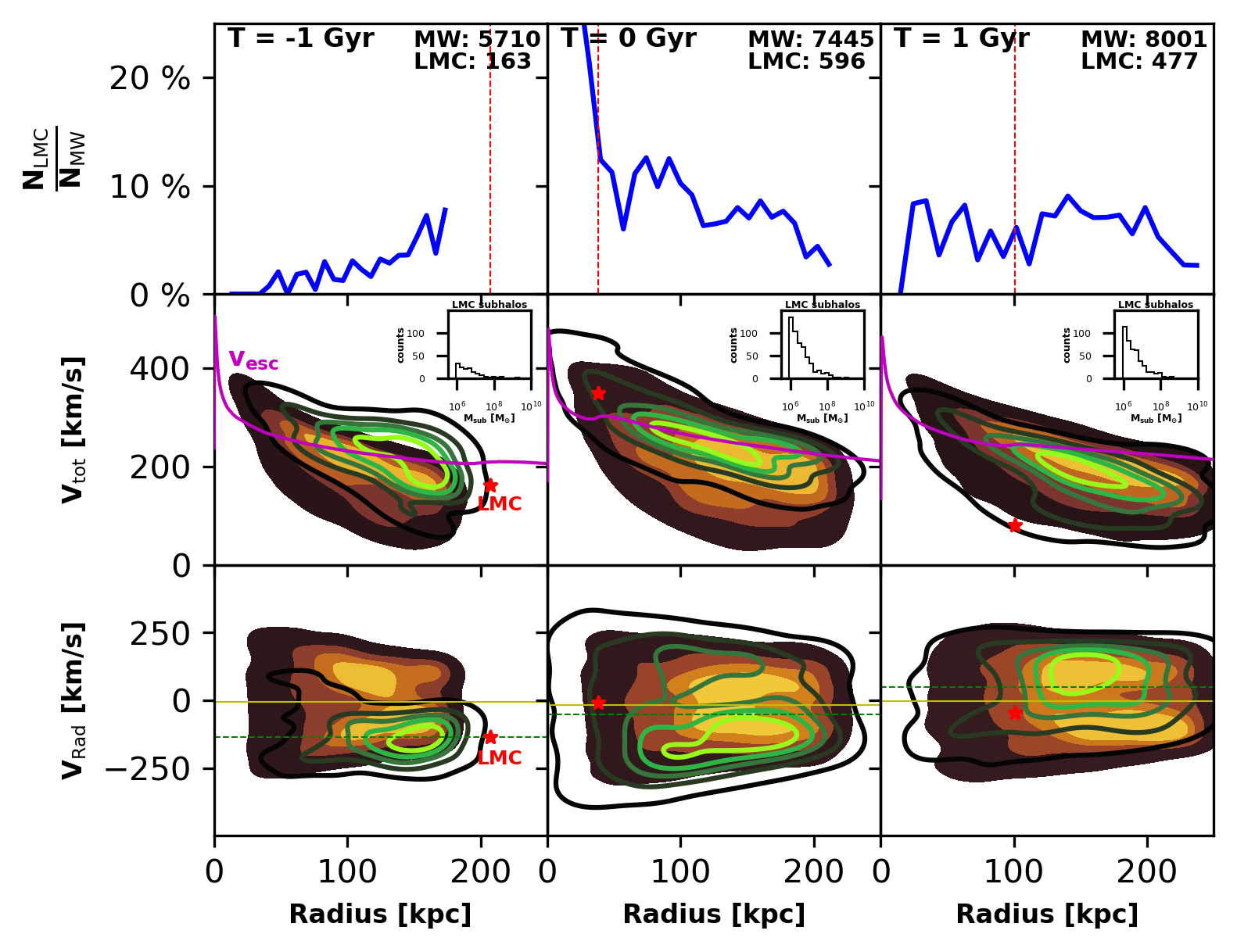}
    \caption{Phase-space evolution of the MW subhalos and the LMC-analog contributed subhalos within the MW's virial radius. Each column shows the distributions at different times: --1 Gyr (left), 0 Gyr (center), and 1 Gyr (right). Top row: fraction of the contributed subhalos as a function of distance. Only a small fraction of subhalos (less than 10\%) are brought in by the LMC-analog, yet they induce significant changes (around 20\%) in the spherically averaged subhalo number density near $T = 0$ Gyr, especially around the LMC-analog's pericentric distance. The red dashed line shows the location of the LMC-analog in each snapshot. Middle: the total velocity distribution in the Galactocentric coordinates versus radius for the MW background (filled contours) and the contributed (open contours) subhalos, along with the host's escape velocity curve (magenta) at each time. Insets show mass distribution of the contributed subhalos. The tangential velocity distribution of the subhalos in the Galactocentric coordinates show similar trends to the total velocity distribution. Bottom: the radial velocity in the Galactocentric coordinates versus radius for both subhalo populations. The mean radial velocity in the Galactocentric coordinates of the MW subhalos and contributed subhalos is marked by solid yellow and dashed green lines respectively. The red star marks the LMC-analog's location. Both the phase-spaces stabilize after the first pericenter. By $T=1$ Gyr, the contributed subhalos distribution matches the MW background as the subhalos undergo phase-mixing.}
    \label{fig:sat_char_m12b}
\end{figure*}

We begin by characterizing the phase-space distribution of both the host subhalos (those within the virial radius of the MW and not brought in by the LMC-analog) and the subhalos brought in by the LMC-analog, hereafter the \emph{contributed} subhalos. We show how the presence of these subhalos influences the overall phase-space distribution and dynamics of subhalos the MW subhalos.

Fig.~\ref{fig:sat_char_m12b} shows the phase-space distribution
at three equally spaced time steps: 1 Gyr before pericenter (left column), when the LMC-analog is at pericenter (center column), and 1 Gyr after pericenter (right column). 
The ratio of contributed subhalos to the total subhalos as a function of radius is plotted in the top row. Despite the contributed subhalos comprising only a fraction (less than 10\%) of the host subhalos, they can increase the spherically averaged subhalo density near the LMC-analog's radial distance (indicated by the red dashed line) by up to 20\%. 

The middle and bottom rows show the total velocity and the radial velocity distributions in the Galactocentric coordinates of the contributed subhalos (depicted as open contours), relative to the background MW subhalo distribution (depicted as filled contours) within the virial radius of the MW. {Notably, both the distributions of the contributed subhalos within the virial radius reveal a distinct leading arm (higher density closer to the host center) compared to the LMC-analog (indicated by the red star) and exhibit different phase-space characteristics compared to the MW. As the merger progresses, the surviving subhalos gradually phase-mix into new orbits resembling those of the MW subhalos, which becomes evident as early as 1 Gyr after the first pericentric passage, aligning with findings in \cite{sales2016identifying} and \cite{barry2023dark}. The tangential velocity distribution of the subhalos (not shown in the Figure) in the Galactocentric coordinates mirrors the trends seen in the total velocity distribution. The escape velocity curves are overlaid in the middle row at each time step. Most of the subhalos are below the escape velocity, however, some exhibit higher total velocities. Many of these subhalos are falling towards the galactic center and thus experience tidal stripping. The insets in the top right corner of the middle row show the mass distribution of the contributed subhalos within the virial radius.}

It's important to highlight that the LMC-analog's position is not centered within the open contour distribution. This is because we are solely analyzing the subhalos situated within the MW's virial radius.
    
Furthermore, At $T=-1$ Gyr, the mean radial velocity of the contributed subhalos is  $\approx$150 km s$^{-1}$ (marked by the green dashed line), corresponding to the analog's radial velocity. By $T=0$ Gyr, the mean radial velocity of the surviving contributed subhalos aligns within $\pm25$ km s$^{-1}$ of the MW subhalo population as they experience phase-mixing.

{In summary, the presence of contributed subhalos can increase the spherically averaged subhalo density by up to 10-40\% (appendix.~\ref{app:m12c_enc}), \cite{barry2023dark} notes a higher boost due to the contributed subhalos in other LMC-analogs. This effect will become more pronounced when considering the azimuthal dependence of the subhalo density distribution. Surviving LMC-analog subhalos, gradually phase-mix into orbits resembling the MW subhalos. However, before $T = 0$ Gyr, the contributed subhalos can shift the mean radial velocity of subhalos, thus affecting the stream--subhalo encounter rates (eq.~\ref{eq:dNenc_dt}). These effects are most prominent during the first pericentric passage and can either increase or decrease the encounter rates.}   

\subsection{ Temporal and radial evolution of the subhalos velocity distribution}\label{sec:velocity_changes}


\begin{figure*}
    \includegraphics[width=\linewidth]{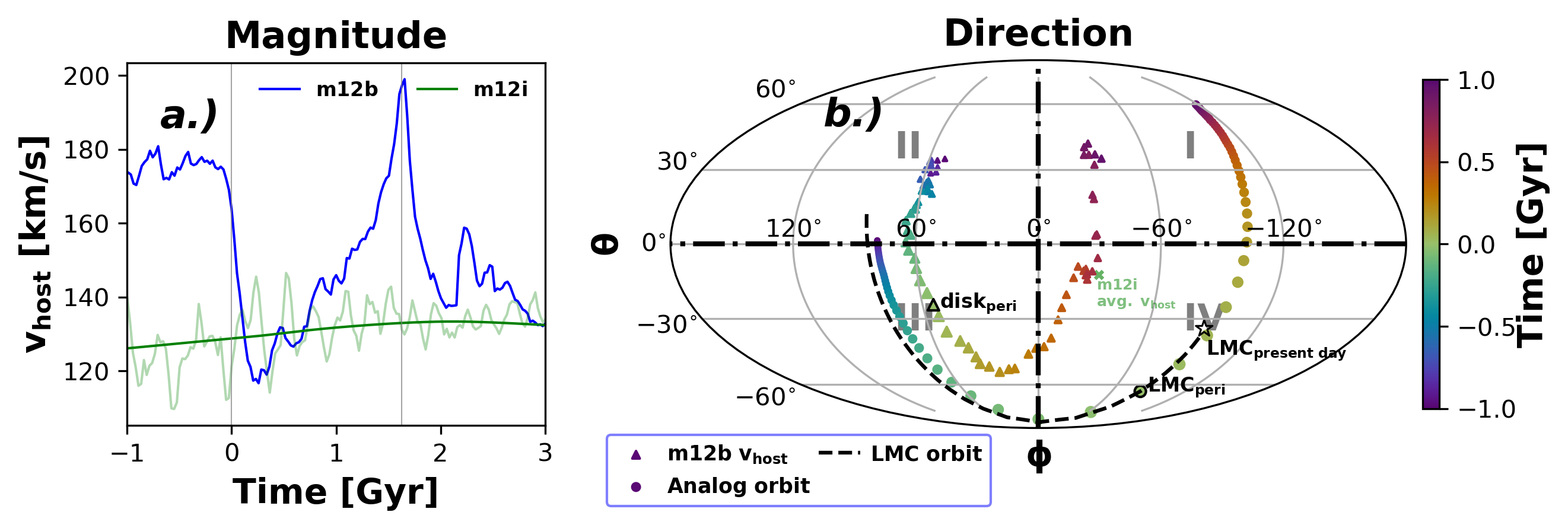}
    \caption{\textbf{a.)} {Center of mass velocity} of the \mb{}  (blue) and \mi{} (green) hosts as a function of time.  The rapid change of 60 km s$^{-1}$ in the host velocity of \mb{} is evident as the LMC-analog passes its first pericenter position at $T = 0$ Gyr (indicated by the gray line), leading to another spike at $T = 1.6$ Gyr (the second pericenter passage). Eventually, the effect stabilizes as the LMC-analog undergoes tidal stripping around the MW's potential. No such changes are noted in \mi{}. \textbf{b.)} The triangles mark the unit vector of the relative central velocity in spherical coordinates quantifying the directional response of the MW center to the LMC-analog orbit marked with circles for \mb{} as a function of time (color bar). The size of the markers is relative to the proximity of the LMC-analog to the MW center. {The MW center moves towards the direction of the LMC-analog}, with the maximum degree of dislodgement occurring at $T = 0$ Gyr. We only show the average direction of the MW center response (marked with a green cross) for \mi{} as no distinct trends are observed. The dashed line shows the orbit of LMC from \citet{garavito2019hunting} up to the present day.}
    \label{fig:reflex_mot}
\end{figure*}

The local subhalo density and velocity distributions are not only affected by the contribution of additional subhalos from the LMC-analog. They are also perturbed by the shift of the COM frame induced by the presence of the LMC-analog. {To illustrate this, we show in} Fig.~\ref{fig:reflex_mot} a.) the center of mass velocity of the MW's host over time for both \mb{} (blue) and \mi{} (green) in the reference frame of the cosmological volume. The MW center of mass is determined using the shrinking spheres method described in \citep{power2003inner}. The total central velocity is calculated using the mean velocities of stars within 10~kpc of the MW's center of mass. 

As the LMC-analog approaches its first pericentric passage, there is a 60 km s$^{-1}$ change in the MW central velocity primarily caused by the LMC-analog moving the COM of the system away from the host's center. This persists for 1--1.5 Gyr after the first pericenter as the LMC-analog and the MW orbit their common COM. The effect reaches another peak as the system approaches its second pericenter passage at approximately 1.7 Gyr. In contrast, such motion is not observed in \mi{} which has a relatively constant velocity of about 130 km s$^{-1}$ in time with deviations of order 10 km s$^{-1}$ in consistency with the results from \cite{salomon2023exploring}.

In Fig.~\ref{fig:reflex_mot} b.), we present the direction of unit vectors $v_\textrm{host}$ of the host with respect to its present-day velocity in spherical coordinates (indicated by triangle markers). We also show the trajectory of the LMC-analog (represented by a solid circle marker) as a function of time (color bar) for \mb{}. Each point on the plot indicates the direction and magnitude of the net velocity of the host. The size of the markers is inversely proportional to the galactocentric distance of the LMC-analog. {The LMC orbit from \citet{garavito2019hunting} is shown with the black dashed line up to the present-day location.}

As the LMC-analog transitions from the Q.III to the Q.IV and moves closer to the galactic center, the MW center of mass moves towards the trajectory of the LMC-analog. This effect is most pronounced when the LMC-analog approaches its first pericenter at $T = 0$ Gyr. By the time of the LMC-analog's second pericenter, the induced reflex motion is not as strong as the halo starts to relax and the satellite loses energy. In contrast, \mi{} does not show a systematic trend with the merging satellite and exhibits a random distribution of velocity vectors. The green cross marker represents the average central displacement over the duration of the satellite merger.      
   

\begin{figure}
    \includegraphics[width=\linewidth]{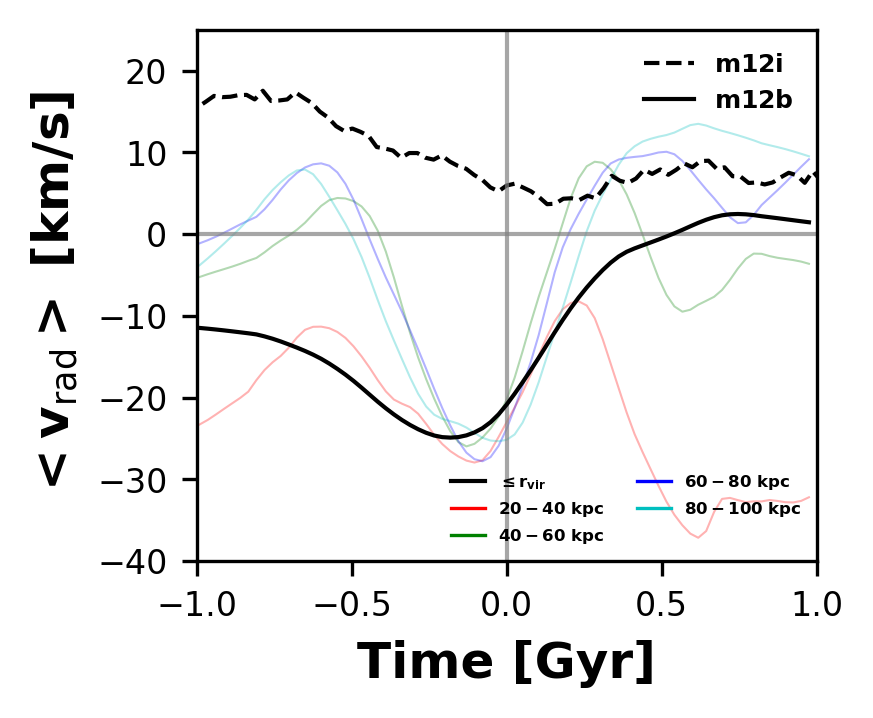}
    \caption{The mean radial velocity of all the subhalos in the Galactocentric coordinates for \mb{} (solid lines) and \mi{} (black dashed line) as a function of time, in different radial bins color-coded. The data are smoothed using the {loess} smoothing method \citep{cleveland1988locally}. The mean radial velocity undergoes significant changes in \mb{} as the LMC-analog approaches its first pericenter passage. However, the radial velocity stabilizes again as the LMC-analog undergoes tidal stripping. In contrast, \mi{} exhibits a relatively constant mean radial velocity throughout the time frame (computed within $\mathrm{r_{vir}}$).}
    \label{fig:mu_glob}
\end{figure}

{We now explore how the velocity of the center-of-mass of the host changes the radial velocity of the subhalos as a function of time.} In Fig.~\ref{fig:mu_glob}, we plot the mean radial velocity of subhalos in Galactocentric coordinates $<v_\textrm{rad}>$ in multiple 20 kpc radial bins. We observe {variations in the mean radial velocities} at the LMC-analog's first pericenter {(T=0) and in subsequent apocenters and pericenter} in \mb{} for all the subhalos within the virial radius (solid black). These variations are both due to the motion of the center of mass of the host and the contribution from the LMC-analog subhalos population with its own distinct phase-space properties. Similarly, the mean tangential velocity of subhalos in the Galactocentric coordinates also changes by about 30 km s$^{-1}$.  
The most significant deviations are observed in the 40--60 kpc bin (solid green) and the 60--80 kpc bin (solid blue), corresponding to the distances probed by the LMC-analog during the given time interval. The high variations in 20--40 kpc bin are attributed to a small number of statistics. \mi{} (dashed line) demonstrates a relatively low and constant mean value. {Interestingly, the velocity dispersion (not shown here) remains relatively constant azimuthally at a radius and in time in both \mi{} and \mb{}.}


\begin{figure}
    \includegraphics[width=\linewidth]{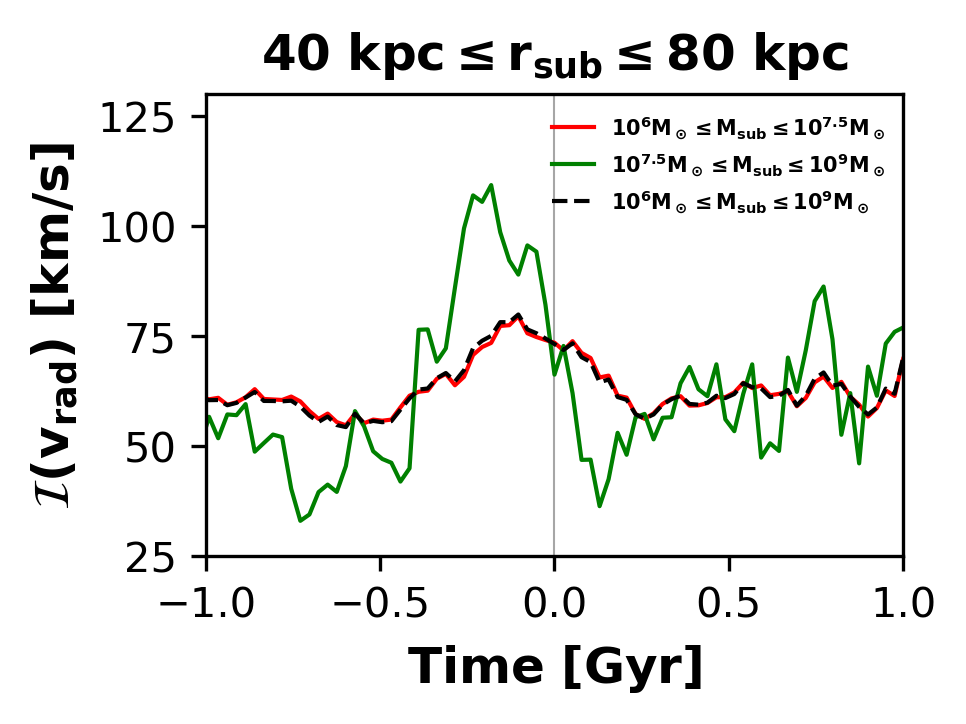}
    \caption{{The probability distribution integral from eq.~\ref{eq:I_vR} as a function of time computed using the 3D radial velocity distribution (${v}_\textrm{rad}$) in galactocentric coordinates of subhalos in low mass ($10^{6}-10^{7.5}$ \Msol{}, red) and high mass ($10^{7.5}-10^9$ \Msol{}, green) bins, and all the subhalos (black) orbiting at a distance of 40-80 kpc from the galactic center. $\mathcal{I}({v}_\textrm{rad})$ for different mass bins show similar variation and behavior. An overall increase in $\mathcal{I}({v}_\textrm{rad})$ is noted right before the first pericenteric passage at 0 Gyr. The high jitter observed for the $10^{7.5}-10^9$ \Msol{} mass bin arises from small number statistics, which affects the convergence of the integral due to the low number of subhalos in this high mass bin.}   
    \label{fig:boost_evolv_mass}}
\end{figure}

The changes in the velocity distribution of the subhalos also affect the first moment of the velocity distribution $\mathcal{I}({v}_\textrm{rad})$ in the galactocentric coordiantes.  Here we quantify those changes as a function of time. Fig.~\ref{fig:boost_evolv_mass} shows the temporal evolution of  ($\mathcal{I}({v}_\textrm{rad})$) computed numerically using eq.~\ref{eq:I_vR}  for subhalos in different mass bins (color-coded). These subhalos orbit within a distance range of 40-80 kpc from the galactic center. The $\mathcal{I}({v}_\textrm{rad})$ for subhalos in different mass ranges exhibits similar trends over time (vary between 50-100 km s$^{-1}$), yet the amplitudes are different by 25\%. Notably, there is an increase in $\mathcal{I}({v}_\textrm{rad})$ right before the LMC-analog's first pericentric passage at 0 Gyr as a result of the COM motion of the host. The black curve, which accounts for all subhalos, closely matches the lower mass bin curves, reinforcing that the overall trend is consistent. The high jitter observed in the $10^{7.5}-10^9$ \Msol{} mass range results from small number statistics, which causes fluctuations in the integral and affects its convergence due to the lower number of subhalos in this mass bin. The consistency in $\mathcal{I}({v}_\textrm{rad})$ highlights that the overall velocity field is mostly decoupled to the mass and density field of subhalos. While we have only considered spherically averaged effects here, this decoupling between the velocity field and the mass of the subhalos is expected to persist even in spatially localized regions. 

\subsection{distance and azimuthal dependence of \texorpdfstring{$\mu$}{mu} and \texorpdfstring{$\sigma$}{sigma}}\label{sec:radial_dependence}
       

\begin{figure}
    \includegraphics[width=\linewidth]{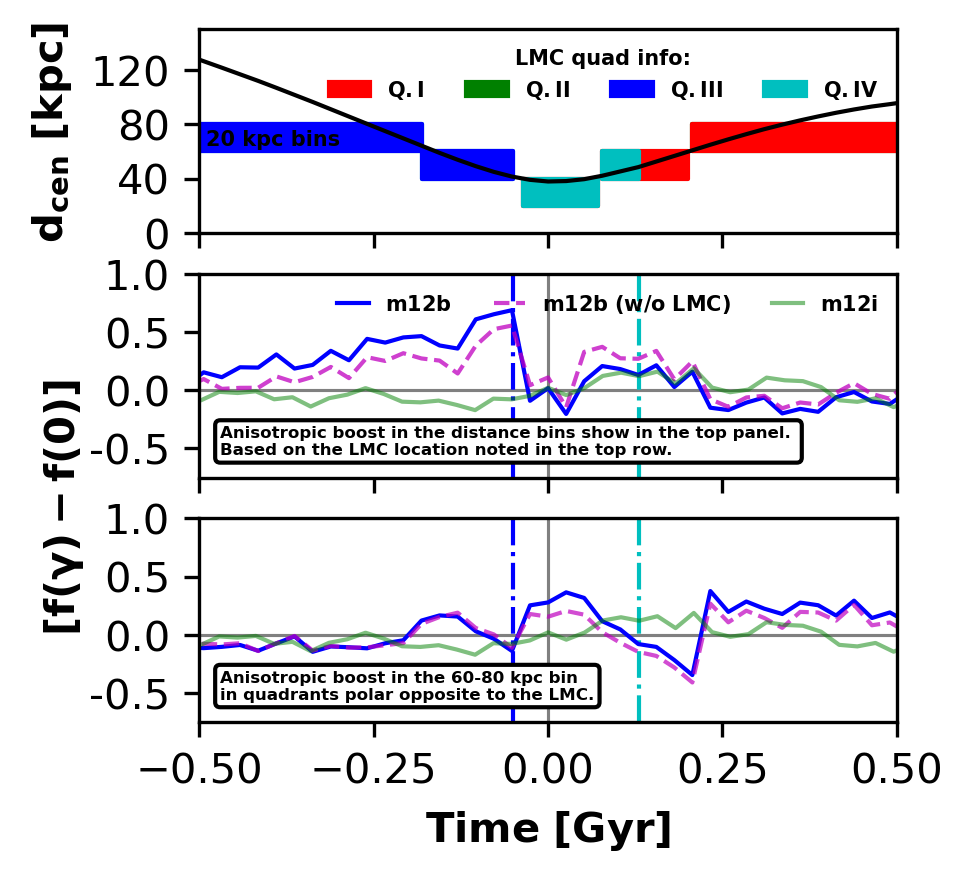}
    \caption{The top row plots the distance from the center of the LMC-analog as a function of time along with the distance bins of width 20 kpc in different quadrants (color-coded) as shaded regions. The middle row plots the change in anisotropic boost factor $\Big[f(\gamma) - f(0)\Big]$ (refer to eq.\ref{eq:num_gauss}) (in Galacocentric coordinates) as a function of time for \mb{} with contributed subhalos (dark blue), \mb{} without contributed subhalos (magenta), and \mi{} (green) evaluated in the distance bins and quadrants marked in the top row (middle row) and in the 60-80 kpc bin in the quadrants polar opposite to the LMC (bottom row). Both cases with and without contributed subhalos exhibit up to a 60\% increase in the boost factor around the LMC-analog, with peak effect just before $T=0$ Gyr. The outermost regions opposite to the LMC-analog also experience a higher boost of about 40\% after the first passage. \mi{}, the galaxy with no massive satellites, shows a variation of approximately 10\% attributed to halo evolution. These values can be attributed to computations corresponding to a stream assumed to be in a fully circular orbit at the center of each distance bin in the respective quadrant.}
    \label{fig:Giant_mess}
\end{figure}

As we have demonstrated in Section~\ref{sec:phase-space} the location of the contributed subhalos from the LMC-analog  varies  as the satellite orbits the host.  Moreover, due to the host's COM motion, the radial velocities of the entire population of subhalos change as a function of radius and time. Here we further explore the angular variations that the radial velocity distribution induces in the $\mathcal{I}({v}_\textrm{rad})$. To simplify our calculation we assume that the radial velocity distribution around the streams is Gaussian. We found this approximation to be accurate at the 10\% level as shown in Appendix~\ref{app:gaussianity}.  We however do not assume that the mean velocity of the radial velocity distribution $\tilde{\mu}$ is zero. The Gaussian approximation of the encounter rates, as defined in eq.~\ref{eq:num_gauss}. To quantify the degree of asymmetry in different regions of the sky we compare our results to the case of $\tilde{\mu}=0$.

In Fig.~\ref{fig:Giant_mess}, the top panel plots the LMC-analog distance from the center as a function of time with the LMC-analog's quadrant marked with shaded color regions at each time step. We also show selected 20 kpc bins between 20-80 kpc interval based on the LMC-analog's location. In the middle panel, we plot the change in the anisotropic boost factor, that is $[f(\gamma) - f(0)]$ as a function of time in the selected 20 kpc bins (show in the top panel) for \mb{} with contributed subhalos (labeled as \mb{}, dark blue) and without contributed subhalos (labeled as \mb{} (w/o LMC subhalos), magenta), and \mi{} (green), the galaxy without massive satellites. The bottom panel shows the change in the anisotropic boost factor within the 60-80 kpc bin in the quadrant polar opposite to the LMC-analog at each time step. Also, these calculations act as proxies for streams in a fully circular orbit centered in each bin in the respective quadrants.

The anisotropic boost factor in both \mb{} with and without contributed subhalos increases up to 60\% as the LMC-analog approaches its first pericenter. While both cases for \mb{} show a positive increase (a positive net inflow of subhalos in the bin), \mb{} with contributed subhalos trends are consistently higher than \mb{} without contributed subhalos by approximately 20\%. The trend increase in \mb{} without the contributed subhalos is caused by the analog's effect on the velocity distribution parameters locally. As the contributed subhalos begin to experience tidal stripping, the trends in both cases for \mb{} converge at around 0.1 Gyr. We note a boost of about 40\% right after the first pericentric passage, majorly in Q. II in the outer bin due to the reflex motion as the LMC-analog moves from Q. III to Q. IV. For \mi{} (green), the variation in boost factor stays within $\pm 10\%$, primarily influenced by the halo evolution and the observed anisotropic velocity distribution \citep{cunningham2019halo7d}.

\subsection{Impact on the local subhalo density}\label{sec:subhalos_density}

\begin{figure*}
    \includegraphics[width=\linewidth]{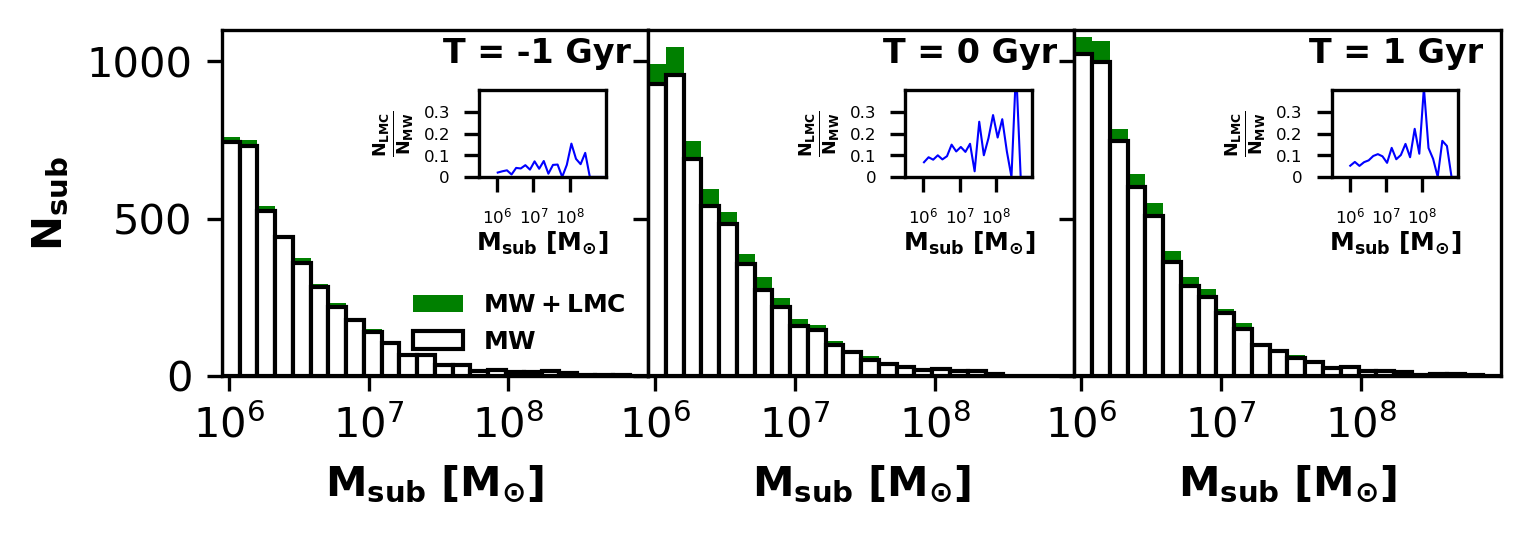}
    \caption{{The total number of subhalos in each decade of mass for the MW subhalos black and MW subhalos combined with those contributed by the LMC-analog (green) that are within the virial radius of the MW at different times: –1 Gyr (left), 0 Gyr (center), and 1 Gyr (right). The insets in each column plot the fraction of LMC to MW subhalos in different mass bins. Around 0 Gyr, The fraction of contributed subhalos are roughly similar (10-20\%) across bins of mass $\leq 10^8$ \Msol{} (dark subhalos). Higher fractions of subhalos with masses  $\geq 10^8$ \Msol{} arise from small number statistics. The contributed subhalo fractions are higher in spatially localized regions (see Fig.~\ref{fig:sat_char_m12b} and Fig.~\ref{fig:subhal_quad_count}).}}
    \label{fig:subhal_mass_fx}
\end{figure*}

\begin{figure*}
    \includegraphics[width=\linewidth]{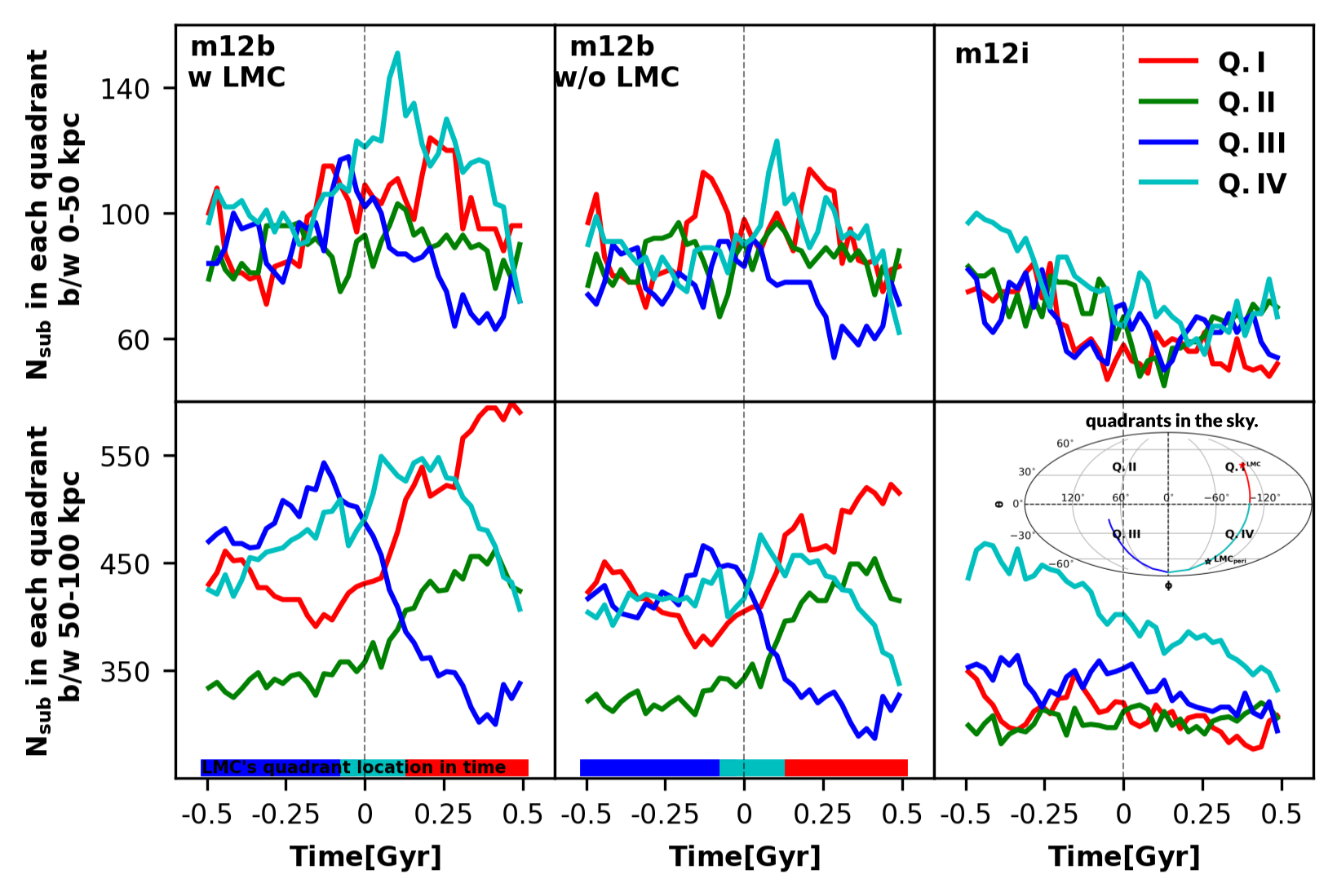}
    \caption{The number of subhalos in each quadrant (defined in Table~\ref{tab:quad_sky}) marked with different colors as a function of time within a time interval of $\pm 0.5$ Gyr around the first pericenter of the LMC-analog. The top row shows the subhalos in the inner halo (0--50 kpc), while the bottom row shows the subhalos in the outer halo (50--100 kpc). In \mb{}, we see large variations in the number of subhalos as a function of time. For example, prior to pericenter (--0.5 Gyr onwards), when the LMC-analog is moving from Q. III to Q. IV, the number of subhalos sharply peaks in Q. II (opposite quadrant) in the middle panel. Looking at the left panel, in Q. IV the number of subhalos increases after pericenter where the LMC-analog moves to, while in Q. III the number of subhalos decreases after pericenter as the LMC-analog leaves Q. III. These changes are the results of the combination of the DM halo response (DM wake and collective response). The lower bar in \mb{} panels mark the analog's quadrant information with color in time.}
    \label{fig:subhal_quad_count}
\end{figure*}

The LMC-analog perturbs the number density $n_\textrm{sub}$ of subhalos in two distinct ways. 

Firstly, it introduces an anisotropically distributed subhalo population that is dragged along by the satellite's motion. Secondly, it induces a response in the MW's DM halo. As the LMC approaches its first pericentric passage, the inner galaxy (consisting of the disk and halo system within 30 kpc) reacts more rapidly than the outer galaxy, resulting in a relative displacement between the two reference frames: the inner and outer galaxy frames. This displacement gives rise to a north-south asymmetry in density, which is referred to as the collective response \citep{garavito2021quantifying, salomon2023exploring}. The collective response includes a large-amplitude weakly damped dipole mode, as well as several other low-amplitude modes \citep{weinberg2023new}. Additionally, the satellite induces a trailing dynamical friction wake in the DM halo \citep{garavito2019hunting}. 

Fig.~\ref{fig:subhal_mass_fx} plots the total number of subhalos within the virial radius of the MW in decades of total subhalo mass for the MW subhalos (black) and MW subhalos combined with LMC contributed subhalos (green) at equally spaced times: -1 Gyr (left), 0 Gyr (center), 1 Gyr (right). The insets in each panel plots the fraction of LMC to MW subhalos as a function of each mass decade and is found to be fairly consistent (10-20\%) around 0 Gyr. This indicates that the subhalos contributed by the LMC-analog do not have significantly different masses compared to the MW subhalos.
                
Fig.~\ref{fig:subhal_quad_count} illustrates the variation in the number of subhalos over time in each quadrant for the inner halo (0-50 kpc, top row) and the outer halo (50-100 kpc, bottom row) in the case of \mb{} with and without contributed subhalos (left and middle column, respectively), as well as for \mi{} (right column). {The color bar at the bottom represents the quadrant location of the LMC-analog in \mb{} during a time interval of $\pm 0.5$ Gyr.} 

For \mi{}, the number of subhalos remains relatively consistent across all quadrants over time, except for Quadrant IV in the outer halo, where the systematically increased subhalo population is caused by the presence of an orbiting dwarf galaxy. 

In the case of \mb{}, we identify the specific effects that contribute to perturbations in the number of subhalos within a time interval of approximately $\pm 0.5$ Gyr. Q.I has the major contributions from a combination of collective response and leading contributed subhalos.

\begin{itemize}
    \item \textbf{contributed subhalos}: We observe a relative increase in the number of subhalos in the outer halo Q. III (between -0.3 to -0.05 Gyr), Q. IV (between -0.05 to 0.13 Gyr), and Q. I (0.13 Gyr onwards) when comparing the left and middle columns. Additionally, there are enhancements in Q. IV of the inner halo at 0 Gyr, corresponding to the satellite being at a pericenter distance of 38 kpc. 
    
    \item \textbf{Collective response}: A significant overdensity is observed in the outer halo (middle column), primarily in Q. I and II, as the satellite approaches pericenter in Q. IV (starting from approximately -0.5 Gyr).  

    \item \textbf{DM dynamical friction wake}: The middle column shows a consistent number of subhalos after the LMC-analog leaves a specific quadrant. In the outer halo, the subhalos predominantly trail the LMC-analog in Q. III (near 0 Gyr) and Q. IV (between 0.12 to 0.25 Gyr).

\end{itemize}

These effects collectively contribute to the observed perturbations in the number density of subhalos and their respective quadrants within the specified time interval for \mb{}.
\section{Encounter rates for simulated and real streams}\label{sec:results}
In this section, we present our results for the stream--subhalo encounter rates for the synthetic streams (Sec.~\ref{sec:enc_rat_synthetic}) and the MW streams (Sec.~\ref{sec:enc_rat_real}). {These encounter rates are calculated using eq.\ref{eq:dNenc_dt} for each time step along the integrated stream orbits computed by directly integrating eq.\ref{eq:I_vR} without applying any Gaussian assumptions}. We subsequently compute the average encounter rates per Gyr using eq.\ref{eq:D_N/D_T}. {We report the number of encounters per Gyr per stream length $\ell_s$ per maximum impact parameter $b_\textrm{max}$ denoted as} $\frac{N_{\mathrm{enc}}}{\mathrm{Gyr}}$ with $\tilde\mu \neq0$\footnote{The notation $\tilde\mu \neq 0$ is used here, but the rates are computed by directly integrating eq.~\ref{eq:I_vR} without applying any Gaussian assumption.}. We also utilize the more traditional approach involving setting $\tilde\mu=0$, denoted as encounter rates with $\tilde\mu = 0$, for the anisotropic boost factor as described in eq.~\ref{eq:num_gauss}. For \mb{}, we further compute the encounter rates by excluding the contributed subhalos, denoted as encounter rates without LMC. Our simulations are limited by the subhalo resolution and the presence of artificial disruption near the LMC-analog \citep{van2018disruption, green2021tidal}, which might lead to a systematic underestimation of the subhalo number densities, and therefore the encounter rates. \citet{barry2023dark} showed that increasing the particle resolution by a factor of 8 did not affect the number density of subhalos significantly.  

While impact-weighting the encounter rates--where the contribution of individual encounters are weighted by their expected effects on a stream's morphology--is crucial for predicting observable changes in a stream's structure \citep{erkal2015forensics, erkal2016number}, here we focus solely on the increase in encounters due to the presence of the LMC's subhalos and the MW's response to the LMC. We have shown that $\mathcal{I}({v}_\textrm{rad})$ is similar across different mass bins up to a degree of non-convergence in higher mass bins (Fig.~\ref{fig:boost_evolv_mass}), indicating that the velocity field is not correlated with the subhalo masses. Additionally, we demonstrated that the LMC-analog contributes a consistent fraction of subhalos across all mass decades globally while the localized contributions can be significantly higher (Fig.~\ref{fig:subhal_mass_fx}). This highlights that the impact weighting in the presence of the LMC subhalos on a stream morphology would be largely similar to the weighting in the absence of the LMC subhalos.

Moreover, this consistency indicates that our procedure of combining encounter rates across all mass decades is robust. In addition, it mitigates high jitter in the $\mathcal{I}(\tilde{v}_\textrm{rad})$ integral, and given the uniform fractional contribution of subhalos from the LMC-analog, it is straightforward to scale these results with $n_\textrm{sub}$ for any given mass decade \citep{barry2023dark}. This approach ensures that our analysis remains valid and reliable, regardless of the specific mass distribution of the subhalos.

\subsection{Encounter rates of synthetic streams} \label{sec:enc_rat_synthetic}
We present statistically averaged encounter rates using the 5000 synthetic streams introduced in Sec.~\ref{Sec:injected_st} as a function of three main properties. 1) galactocentric distance; 2) location on the sky, defined by the quadrant; 3) angular momentum direction relative to the LMC-analog's angular momentum in \mb{}. 

\subsubsection{Dependence on the pericenter distance and sky location}


\begin{figure}
    \includegraphics[width=\linewidth]{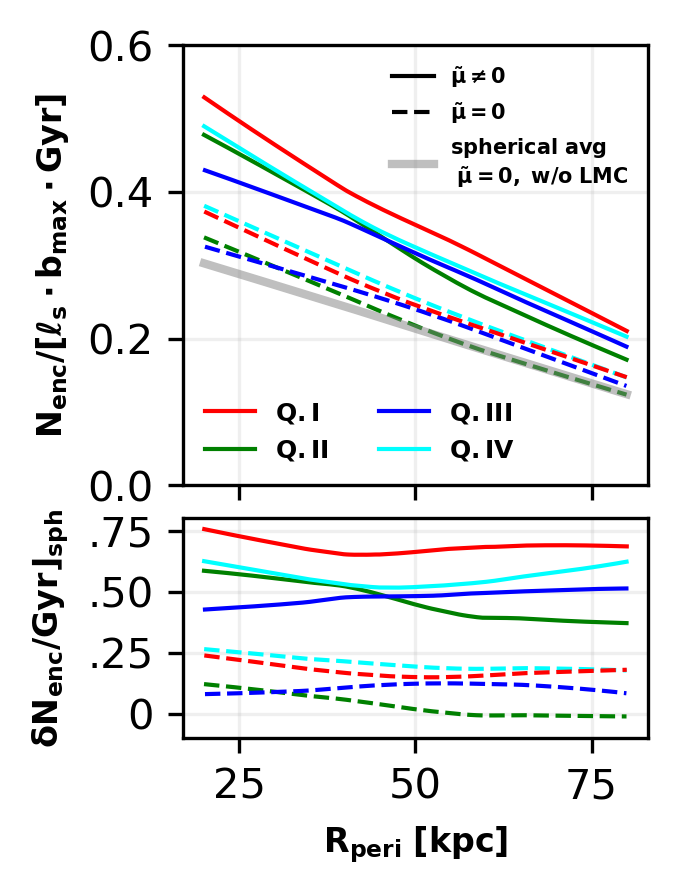}
    \caption{Estimated number of encounters per Gyr {per stream length $\ell_s$  and impact parameter $\mathrm{b_{max}}$} in \mb{} with $\tilde\mu \neq 0$ (solid) and  $\tilde\mu = 0$ (dashed). The rates are plotted as a function of a stream's pericenter distance at $T=0$ Gyr in each quadrant (marked by different colors). The gray curve marks the encounter rates for \mb{} $\tilde\mu = 0$ and without the contributed subhalos, as well as for \mi{} (re-scaled to match the \mb{} halo mass), averaged over all quadrants. These curves are estimated using a lowess regression \citep{cleveland1988locally} from the 5000 synthetic streams. Streams in Q. I and IV have the highest probability of encounters, as the LMC-analog's' position in \mb{} at $T=0$ Gyr leads to a higher concentration of subhalos in Q. IV (at large distances), and the majority of the reflex motion and collective response effects are dominant in Q. I. Outer regions of Q.II has the lowest encounter rates of all, consistent with the spherically averaged rates marked by the gray curve. The second row plots the fractional changes in each encounter rate curve with respect to the spherically averaged rates.}
    \label{fig:enc_rat_m12b_mu_change}
\end{figure}

The encounter rates depend linearly on the number density of subhalos $n_\textrm{sub}$ and on the velocity dispersion of subhalos in the stream-centric coordinates, both of which decrease as a function of the galactocentric distance. \citet{barry2023dark} showed that the number density of subhalos is constant as a function of distance from the center out to 50 kpc at present day, however, we found a decreasing trend in the number density around our integration time $\approx$ 5 Gyr before the present day. The top panel in Fig.~\ref{fig:enc_rat_m12b_mu_change} shows the {encounter rate per Gyr per stream length $\ell_s$  and impact parameter $\mathrm{b_{max}}$} smoothed using the lowess regression \citep{cleveland1988locally} as a function of a stream's pericentric distance in each quadrant (color-coded) defined in Fig.~\ref{fig:LMC_traj_m12b}, along with a spherically averaged encounter rate without the anisotropic boost ($\tilde\mu = 0$) and the contributed subhalos (gray curve). The bottom panel shows the fractional changes in the rates relative to the spherically averaged rates. The overall number of encounters per Gyr ($\mathrm{N_{enc}/Gyr}$) decreases as a function of distance in all quadrants. The $\mathrm{N_{enc}/Gyr}$ values with $\tilde\mu \neq 0$ (solid lines) consistently exhibit enhancements of 25\%-70\%, while the rates with $\tilde\mu = 0$ (dashed lines) have a maximum enhancement of about only 25\% from the changes in the number density of subhalos. 

Q. I and IV curves with $\tilde\mu \neq 0$ show the highest increment in the rates, with increases of up to 50--70\% (bottom panel). The boost in Q. IV can be attributed to {the LMC-analog's contributed subhalo increasing the number density of subhalos and changes in the effective $\tilde\gamma$ (given by $\tilde\mu/\sqrt{2}\tilde\sigma$)}. While Q.I exhibits higher encounter rates due to the collective response and reflex motion. Q. III shows a relatively similar increase at larger distances due to the contributed subhalos and the influence of the DM wake. Q. II has the lowest encounter rates as none of the boosting effects are prominent there. On the other hand, \mi{} (not shown here) has no quadrant dependence and the encounter rates with $\tilde\mu$ are higher by 20\%  when compared with $\tilde\mu=0$. We note that even without an LMC-analog, $\tilde\mu=0$ is generally non-zero, similarly \citet{cunningham2019halo7d} showed that velocity isotropy changes as a function of position on the sky.

\subsubsection{The LMC-analog contribution to the encounter rates}


\begin{figure}
    \includegraphics[width=\linewidth]{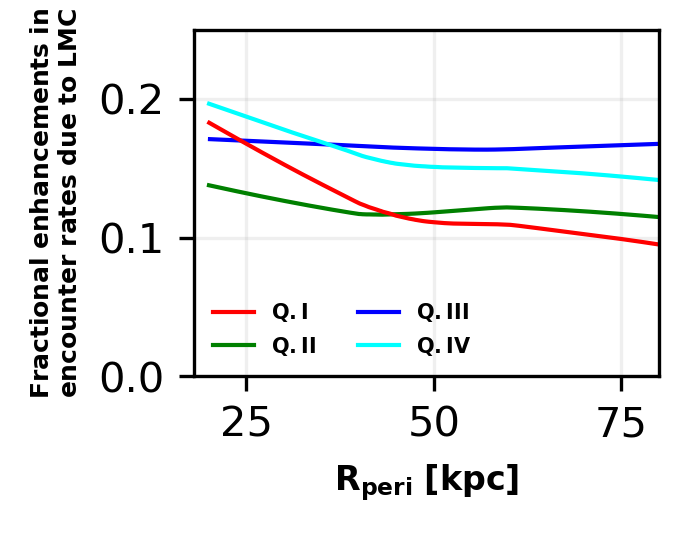}
    \caption{Fractional enhancements in the encounter rates due to the contributed subhalos from the LMC-analog as a function of a stream's pericentric distance in each quadrant {(marked with different colors)} for \mb{}. Streams exhibit a consistent increase of approximately 20\% in encounter rates across all distance ranges in Q.III and inner regions of Q.IV due to the contributed subhalos. In contrast, outer regions of Q.I show a smaller increment of $\sim 10\%$ due to contributed subhalos. Notably, Q.II has a constant increment of about 12\% at larger distances, indicative of the presence of trailing contributed subhalos. These subhalos can boost the encounter rates through the halo by 15\%.}
    \label{fig:enc_rat_diff}
\end{figure}

In Fig.~\ref{fig:enc_rat_diff}, we plot the fractional enhancements in $\mathrm{N_{enc}/Gyr}$ resulting from the presence of LMC-analog's contributed subhalos in \mb{}. These enhancements are computed by fractional change in the encounter rates with $\tilde\mu \neq 0$ in the presence of the contributed subhalos and without these subhalos, using lowess regression. Notably, the trends in fractional enhancements for encounter rates with $\tilde\mu = 0$ show similar patterns.

The contributed subhalos have a more pronounced impact on the encounter rates in Q.III and Q.IV, yielding a substantial increase of nearly 15--40\% (appendix.~\ref{app:m12c_enc}). This observation aligns with the findings from the first row of Fig.~\ref{fig:sat_char_m12b}. In contrast, Q.I and Q.II in the outer halo exhibit relatively lower enhancements, with values not exceeding 12\%. Notably, Q.I in the outer halo showcases the least enhancement, {and therefore the} higher encounter rates seen at larger distances in Q.I in Fig.~\ref{fig:enc_rat_m12b_mu_change} are primarily influenced by the reflex motion effects. We also found a higher localized boost of about 40\% for another LMC-analog system in appendix.~\ref{app:m12c_enc} where the analog falls in closer to the present day \citep{barry2023dark}. 

In summary, our analysis highlights the crucial role of subhalo--stream kinematics, particularly the velocity distribution of subhalos, as a key factor in accurately predicting encounter rates. This results in a minimum overall enhancement of 25\%. Moreover, we observe a notable dependence in the encounter rates on the location in the sky. Q. IV experience an additional enhancement of nearly 10--40\%, driven by the presence of contributed subhalos, while Q. I show increased rates due to the interplay of the collective response and reflex motion. 

\subsubsection{Dependence on the streams' orientation and orbit}


\begin{figure}
    \includegraphics[width=\linewidth]{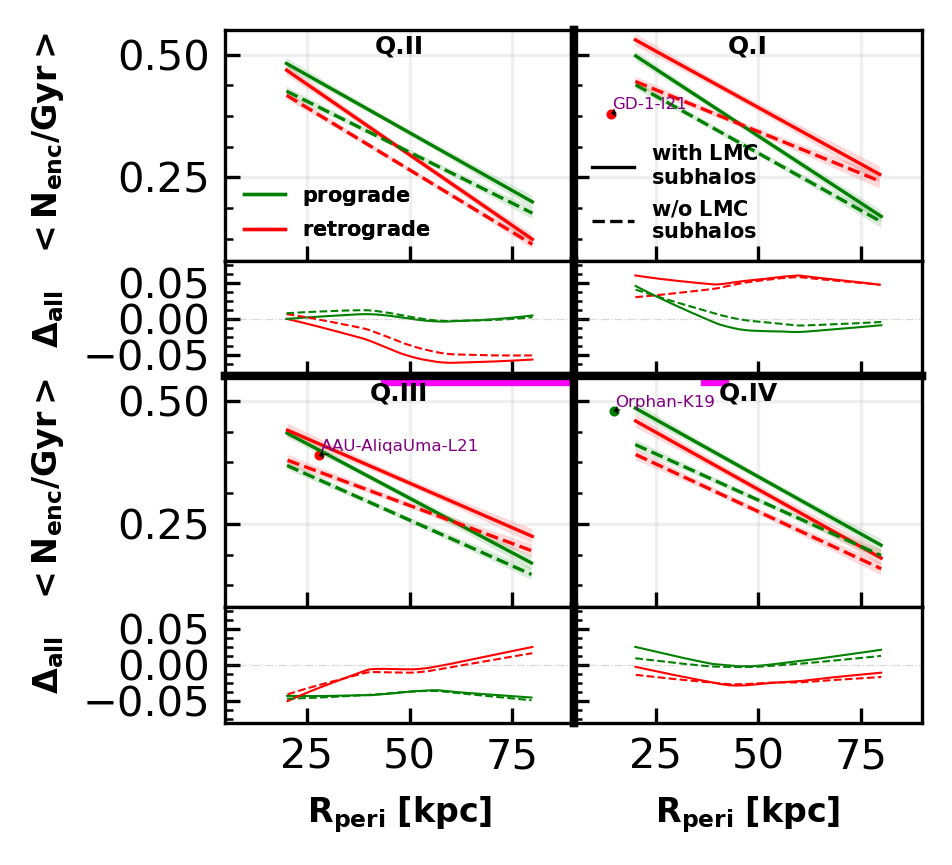}
    \caption{The average encounter rate (row 1 and 3) is estimated using separate robust regressions \citep{robustreg, huber1973robust} for prograde (green) and retrograde (red) streams, taking into account the stream's pericenter distance and sky position (quadrants marked on the panels). These calculations are performed in \mb{} with (solid) and without (dashed) contributed subhalos from the satellite. The difference with respect to the spherically average, denoted as $\Delta_\mathrm{all}$, is also shown in rows 2 and 4 in each quadrant for each curve. The magenta bar marks the LMC-analog's distance in each quadrant within the time frame of the encounter rate calculation. {Additionally, predicted encounter rates for the MW streams GD-1, Orphan, and AliqUma, categorized by their orbit types, are also shown in their respective quadrants, plotted as a function of their present-day distance.} We note that streams in retrograde orbits with respect to the LMC-analog consistently exhibits higher encounter rates in Q.I (regions polar opposite to the analog's trajectory.) }
    \label{fig:enc_rat_reg}
\end{figure}

We categorize the synthetic streams as prograde or retrograde based on whether their orbital motion is in the same or opposite direction to that of the LMC-analog, respectively. Among the 5000 synthetic streams, approximately 50\% exhibit retrograde orbits relative to the LMC-analog's orbit, while the remaining 50\% are in prograde orbits. Figure \ref{fig:enc_rat_reg} plots the averaged encounter rates for streams in prograde (green) and retrograde (red) orbits as a function of a stream's pericentric distance in each quadrant (marked in panels in rows 1 and 3). These rates are computed using robust regression \citep{robustreg, huber1973robust} in \mb{} with (solid lines) and without (dashed lines) contributed subhalos. The sub-panels below each quadrant (row 2 and 4) show the $\Delta_\mathrm{all}$ as a function of a stream's pericentric distance, which is the difference between the average encounter rates in each quadrant for each orbit type and the spherically averaged encounter rate for both orbit types, with and w/o contributed subhalos. The magenta bar in specific panels indicates the distance limits of the LMC-analog within each quadrant during the calculation of encounter rates. 

In Q.I, the retrograde orbits with respect to the LMC-analog consistently exhibit higher encounter rates than prograde orbits (approximately 0.02 - 0.08 more encounters or 50\% higher than the spherically averaged gray curve shown in Fig.~\ref{fig:enc_rat_m12b_mu_change}), regardless of the presence of the contributed subhalos. This is due to the relative bulk motion of subhalos, which aligns in the same direction as the retrograde stream orbits in Q.I, thereby increasing the likelihood of encounters. 

In Q.IV, the prograde orbits in the outer regions have a higher number of encounters (approximately by 0.02 per Gyr) due to their motion aligning with that of the analog which increases the likelihood of encountering subhalos, as the streams spend more time in close proximity to the contributed subhalos. Additionally, the presence of a transient DM wake can further contribute to the increased encounter rates for prograde orbits in the outer regions of Q.III. 

In Q.II, retrograde streams have systematically lower rates than both the spherically averaged rates and prograde orbit rates. We think this may be due to the fact that streams move between quadrants over the full integration time of 0.5 Gyr. The retrograde streams in Q.II preferentially enter via Q.III, where the relative velocity with subhalos from the LMC-analog is high since most subhalos in that quadrant have a bulk velocity in the same sense as the analog's orbit. This then produces an apparent decrement in retrograde encounters in Q.II. Conversely, streams that are moving prograde in 
Q.II at the end of the integration period came from Q.I, which is not strongly affected by the analog's contributed subhalos. This would also produce the difference between the rates calculated with and without including the contributed subhalos that we also observe in Q.II. This quadrant serves as a reminder that the impact of the LMC on stream-subhalo encounters is not quite as simple as we would hope!

Furthermore, the encounter rates with $\tilde\mu = 0$ (not depicted) show no discernible trends between different orbit types across any of the quadrants. However, the introduction of a non-zero $\tilde\mu$ complicates the systematic perspective due to the influence of the stream's apparent motion on the anisotropic boost factor. Additionally, there is no observed dependence or correlation between a stream's orbital eccentricity and encounter rates.

We annotate the GD-1 stream \citep{grillmair2006detection}, the Orphan stream \citep{grillmair2006orphan, belokurov2007orphan, koposov2019piercing}, and the AliqaUma stream \citep{li2021broken} in their respective quadrants and orbit types relative to the LMC analog. The trends observed in our synthetic streams generally align with those observed in the MW streams. The Orphan-Chenab stream closely interacts with the LMC and extends across both hemispheres of the MW \citep{koposov2023s}. Additionally, \cite{lilleengen2023effect} has demonstrated the effects of the deforming MW halo on the proper motion and morphological structure of streams, making it challenging to differentiate between changes caused by subhalo--stream interactions and other influences.

\subsection{Encounter rates of MW streams} \label{sec:enc_rat_real}

We have observed that synthetic streams around halos with an LMC-analog display a diverse range of encounter rates, contingent upon several factors: 1) The specific value of $\tilde\mu$ associated with the stream; 2) the stream location in the sky; and 3) The stream orbital rotation direction in relation to the LMC-analog's orbit. Notably, our findings demonstrated boosted encounter rates for orbits located in the northern hemisphere, and in regions near and diametrically opposite to the LMC.

Expanding upon our analysis, we now proceed to calculate encounter rates per Gyr {per $\ell_s$ and $b_\textrm{max}$} for all the MW streams within our simulation. We employ the same orbital integration methodology as employed for synthetic streams. However, for the MW streams, we substitute the initial conditions with the present-day phase-space coordinates of each stream in the MW. To establish these initial conditions, we utilize the median position and velocity of each stream track. We further adjust these conditions by applying a rotation that emulates the orientation of the actual streams relative to the real LMC as discussed in Sec.~\ref{Sec:injected_st}. 

Our results are summarized in Table~\ref{tab:enc_rate_streams}, we present the computed encounter rates ($\mathrm{N_{enc}/Gyr}$) in both \mb{} and \mi{} for the known MW streams \citep{mateu2023galstreams}. The table also includes the orbital properties of each stream, such as its present-day distance from the host galaxy ($\mathrm{d}$) and the sky quadrant ($\mathrm{Q}$) it occupies at $T = 0$ Gyr in the context of \mb{}. The eccentricity of the orbit ($\mathrm{e}$) and its grade ($\mathrm{G}$) with respect to the LMC-analog in \mb{} are also provided. The grade indicates whether the stream's motion is prograde (in the same direction as the LMC-analog's orbit, denoted by ``+'') or retrograde (opposite to the LMC-analog's orbit, denoted by ``-''). These properties are calculated using the \mb{} potential models of DM halo and baryonic disks from \citet{arora2022stability}. It is crucial to emphasize that around 75\% of the MW streams are situated within a radius of 20~kpc at $T = 0$ Gyr. While interpreting these encounter rates, it's important to be cautious due to the limited resolution in the inner regions. \citet{barry2023dark} showed that the total subhalo population within a 50 kpc radius, 5~Gyr ago, was approximately twice as high for subhalos with $\mathrm{M_{sub}} \geq 10^6$ \Msol{} compared to the present-day population. As a result, we can adjust our encounter rates by dividing them by a factor of 2. For example, \citet{barry2023dark} estimated that the GD-1 stream, with a length of 15~kpc with an average impact parameter of 1.5 kpc, could experience approximately 5 encounters per Gyr using traditional symmetry and isotropy assumptions. With the inclusion of the LMC-induced subhalos, this rate increases twofold. In our analysis, our re-scaled factor for present-day encounter rate for GD-1 with the same stream length and impact parameter is 3.5 encounters per Gyr for \mi{} and 4.3 encounters per Gyr for \mb{}. These numbers are more or less similar because \mi{} has about 1.2 times less subhalos than \mb{} without the LMC-contributed subhalos. Given the location of GD-1 in the sky and distance from the center, we don't expect any boosts in encounter rates due to the LMC, and our numbers are consistent with \citet{barry2023dark} to the first order.      

Among the MW streams analyzed, the Aquarius-W11 stream \citep{williams2011aqua}, Lethe-G09 \citep{grillmair2009Lethe}, and Phlegethon-I21 \citep{ibata2021charting} in the inner halo (within 30~kpc) exhibit the highest encounter rates within the \mb{} simulation. Conversely, the Cetus-Y13 stream \citep{yam2013cetus} has the highest encounter rate in the outer halo (at distances greater than 30 kpc). The streams C-7-I21 \citep{ibata2021charting}, Ophiuchus-C20 \citep{caldwell2020larger}, and OmegaCen-I21 \citep{ibata2021charting} have the lowest encounter rates among the analyzed streams. These streams are located in the inner halo, orbiting within 7~kpc of the galactic center. Additionally, the Eridanus-M17 stream \citep{myeong2017tidal}, positioned at a distance of 100~kpc, exhibits the lowest encounter rates compared to other streams.      

Overall, we observe consistent encounter rate trends for the MW streams across \mi{} and \mb{}, both with and without $\tilde\mu$. Streams situated beyond a present-day distance of 20 kpc in the \mb{} exhibit a clear dependence on location, with enhanced encounter rates in Q. II and IV. Conversely, in \mi{}, streams demonstrate no azimuthal dependence in the encounter rates. Incorporating $\tilde\mu$ yields approximately a 30\% boost for both \mb{} and \mi{}. The sense of rotation of the streams does not notably impact the encounter rates, mainly due to the proximity of these streams to the galactic center, while these effects are prominent for streams positioned in the outer halo regions.   

\begin{ThreePartTable}
    {\tiny
    \begin{longtable}{p{0.15\textwidth}lp{0.025\textwidth}cp{0.025\textwidth}cp{0.025\textwidth}cp{0.025\textwidth}cp{0.025\textwidth}cp{0.01\textwidth}|}
    \caption{Encounter rates per Gyr per unit stream length (kpc) per maximum impact parameter (kpc) of the MW streams in \mb{} and \mi{}.} 
    \label{tab:enc_rate_streams} \\
    \toprule
    \multirow{2}{*}{\textbf{Name}} & \multirow{2}{*}{\textbf{d$^*$ [kpc]}} & \multirow{2}{*}{\textbf{Q$^*$}} & \multirow{2}{*}{\textbf{G}} & \multirow{2}{*}{\textbf{e}}  &  \multicolumn{2}{c}{\textbf{Enc. rates}} \\ 
        &   &   &   &   & \multicolumn{1}{c}{\textbf{m12b}} & \multicolumn{1}{c}{\textbf{m12i}} \\
    \midrule
    \endfirsthead
    \toprule
    \textbf{Name} & \textbf{d$^*$ [kpc]} & \textbf{Q$^*$} &  \textbf{G} &  \textbf{e} &  \multicolumn{2}{c}{\textbf{Enc. rates}} \\ 
    \multicolumn{1}{l}{} & \multicolumn{1}{l}{} & \multicolumn{1}{l}{} & \multicolumn{1}{l}{} & \multicolumn{1}{l}{} & \multicolumn{1}{c}{\textbf{m12b}} & \multicolumn{1}{c}{\textbf{m12i}} \\
    \midrule
    \endhead
    \midrule
    \multicolumn{7}{r}{{Continued on next page}} \\
    \midrule
    \endfoot
    
    \bottomrule
    \endlastfoot
       20.0-1-M18 &  20.0 &        III &           - &    0.3 &              0.4 &              0.4 \\
         300S-F18 &  20.6 &          I &           + &    0.6 &              0.4 &              0.3 \\
 AAU-AliqaUma-L21 &  27.9 &        III &           - &    0.5 &              0.4 &              0.4 \\
    AAU-ATLAS-L21 &  23.1 &        III &           - &    0.5 &              0.4 &              0.4 \\
      Acheron-G09 &   6.1 &         II &           + &    0.2 &              0.3 &              0.3 \\
          ACS-R21 &  18.7 &         II &           + &    0.1 &              0.5 &              0.3 \\
      Alpheus-G13 &   8.0 &        III &           - &    0.1 &              0.2 &              0.3 \\
     \textbf{Aquarius-W11} &   8.1 &        III &           + &    0.6 &              0.8 &              0.4 \\
         C-19-I21 &  20.8 &        III &           + &    0.5 &              0.3 &              0.3 \\
          C-4-I21 &  10.1 &         II &           - &    1.0 &              0.4 &              0.3 \\
          C-5-I21 &  12.1 &        III &           + &    0.8 &              0.4 &              0.2 \\
          C-7-I21 &   4.0 &        III &           + &    0.4 &              0.1 &              0.2 \\
          C-8-I21 &   7.0 &        III &           - &    0.4 &              0.5 &              0.4 \\
          C-9-I21 &  12.1 &        III &           - &    0.1 &              0.4 &              0.3 \\
    Cetus-New-Y21 &  20.8 &        III &           - &    0.3 &              0.4 &              0.3 \\
  Cetus-Palca-T21 &  34.6 &        III &           - &    0.4 &              0.3 &              0.3 \\
        Cetus-Y13 &  35.3 &        III &           - &    0.2 &              0.4 &              0.4 \\
      Cocytos-G09 &   9.6 &          I &           + &    0.2 &              0.4 &              0.4 \\
       Corvus-M18 &  11.0 &          I &           + &    0.5 &              0.5 &              0.3 \\
        Elqui-S19 &  50.1 &        III &           - &    0.5 &              0.3 &              0.4 \\
     Eridanus-M17 & 100.0 &        III &           - &    0.6 &              0.1 &              0.1 \\
       \textbf{Gaia-1-I21} &   8.2 &          I &           - &    0.6 &              0.6 &              0.3 \\
      Gaia-10-I21 &  18.4 &          I &           - &    0.1 &              0.5 &              0.3 \\
      Gaia-11-I21 &  12.0 &         II &           - &    0.9 &              0.4 &              0.4 \\
      Gaia-12-I21 &  18.2 &        III &           + &    0.3 &              0.3 &              0.3 \\
       Gaia-2-I21 &  10.8 &        III &           - &    0.2 &              0.4 &              0.4 \\
       \textbf{Gaia-3-M18} &  12.9 &          I &           + &    0.6 &              0.6 &              0.3 \\
       \textbf{Gaia-4-M18} &  14.5 &          I &           + &    0.5 &              0.6 &              0.2 \\
       Gaia-5-M18 &  25.5 &          I &           + &    0.2 &              0.5 &              0.3 \\
       \textbf{Gaia-6-I21} &  10.4 &          I &           - &    0.2 &              0.5 &              0.2 \\
       Gaia-7-I21 &   8.7 &          I &           + &    0.4 &              0.4 &              0.2 \\
       Gaia-8-I21 &   8.6 &          I &           - &    0.3 &              0.4 &              0.2 \\
       Gaia-9-I21 &   9.4 &         II &           + &    0.9 &              0.4 &              0.4 \\
         GD-1-I21 &  14.2 &          I &           - &    0.1 &              0.4 &              0.3 \\
     Gunnthra-I21 &   6.3 &        III &           + &    0.4 &              0.2 &              0.3 \\
       Hermus-G14 &  17.3 &         II &           + &    0.5 &              0.6 &              0.3 \\
         Hrid-I21 &   8.0 &         II &           - &    0.9 &              0.6 &              0.4 \\
       Hyllus-G14 &  17.4 &         II &           + &    0.5 &              0.6 &              0.4 \\
        Indus-S19 &  13.6 &        III &           + &    0.2 &              0.4 &              0.4 \\
          Jet-F22 &  34.0 &         IV &           - &    0.7 &              0.4 &              0.3 \\
     Jhelum-a-B19 &  11.3 &        III &           + &    0.6 &              0.4 &              0.5 \\
     Jhelum-b-B19 &  11.4 &        III &           + &    0.6 &              0.4 &              0.5 \\
        Kshir-I21 &  14.8 &         II &           - &    0.6 &              0.4 &              0.3 \\
       Kwando-G17 &  22.9 &        III &           - &    0.2 &              0.4 &              0.3 \\
       Leiptr-I21 &  13.7 &         IV &           + &    0.7 &              0.4 &              0.3 \\
        Lethe-G09 &  13.3 &          I &           + &    0.2 &              0.5 &              0.3 \\
         LMS1-Y20 &  12.4 &          I &           + &    0.6 &              0.4 &              0.2 \\
           M2-I21 &   9.8 &        III &           + &    0.9 &              0.3 &              0.3 \\
          M30-S20 &   7.0 &        III &           - &    0.3 &              0.4 &              0.3 \\
           M5-G19$^\dagger$ &  13.7 &          I &           - &    \_ &              0.7 &              0.4 \\
          M68-P19 &   8.9 &          I &           + &    0.4 &              0.4 &              0.4 \\
          M92-I21 &  10.3 &         II &           - &    0.6 &              0.3 &              0.3 \\
     Molonglo-G17 &  21.0 &        III &           - &    0.2 &              0.4 &              0.3 \\
    Monoceros-R21 &  18.2 &         II &           + &    0.1 &              0.4 &              0.3 \\
 Murrumbidgee-G17 &  21.0 &        III &           - &    0.3 &              0.4 &              0.3 \\
      NGC1261-I21 &  21.4 &         IV &           + &    0.9 &              0.5 &              0.4 \\
      NGC1851-I21 &  17.8 &         IV &           + &    0.7 &              0.3 &              0.3 \\
      NGC2298-I21 &  17.4 &         IV &           + &    0.9 &              0.4 &              0.4 \\
       \textbf{NGC288-I21} &  13.7 &        III &           + &    0.6 &              0.6 &              0.2 \\
      NGC3201-P21 &  11.0 &         IV &           + &    0.6 &              0.5 &              0.3 \\
      \textbf{NGC5466-G06} &  17.5 &          I &           + &    0.2 &              0.5 &              0.2 \\
      \textbf{NGC6362-S20} &   5.0 &         IV &           - &    0.7 &              0.6 &              0.2 \\
      NGC6397-I21 &   5.9 &        III &           - &    0.5 &              0.2 &              0.1 \\
     OmegaCen-I21 &   7.0 &         IV &           - &    0.4 &              0.1 &              0.1 \\
    \textbf{Ophiuchus-C20} &   4.3 &          I &           - &    0.8 &              0.2 &              0.0 \\
      Orinoco-G17 &  22.2 &        III &           - &    0.2 &              0.4 &              0.3 \\
       Orphan-K19 &  14.8 &         IV &           + &    0.2 &              0.5 &              0.3 \\
        Pal13-S20 &  24.8 &        III &           - &    0.6 &              0.4 &              0.3 \\
        Pal15-M17 &  31.6 &          I &           + &    0.3 &              0.4 &              0.2 \\
        Pal5-PW19$^\dagger$ &  16.6 &          I &           - &    \_ &              0.7 &              0.6 \\
        Palca-S18 &  37.5 &        III &           - &    0.2 &              0.3 &              0.3 \\
     \textbf{Parallel-W18} &  16.9 &          I &           + &    0.4 &              0.5 &              0.2 \\
      Pegasus-P19 &  18.6 &         II &           - &    0.7 &              0.6 &              0.3 \\
\textbf{Perpendicular-W18} &  17.5 &          I &           + &    0.2 &              0.6 &              0.2 \\
   Phlegethon-I21 &   6.7 &        III &           + &    0.6 &              0.7 &              0.4 \\
      Phoenix-S19 &  18.5 &        III &           - &    0.2 &              0.4 &              0.4 \\
        PS1-A-B16 &  13.6 &        III &           - &    0.1 &              0.4 &              0.3 \\
        \textbf{PS1-B-B16} &  18.7 &         IV &           + &    0.5 &              0.5 &              0.3 \\
        PS1-C-B16 &  17.5 &        III &           - &    0.7 &              0.6 &              0.4 \\
        PS1-D-B16 &  28.1 &         IV &           + &    0.5 &              0.4 &              0.3 \\
        \textbf{PS1-E-B16} &  17.6 &         II &           + &    0.2 &              0.6 &              0.2 \\
         Ravi-S18 &  19.7 &        III &           - &    0.8 &              0.4 &              0.4 \\
  Sagittarius-A20 &   8.3 &        III &           - &    0.9 &              0.3 &              0.4 \\
    Sangarius-G17 &  27.1 &         IV &           + &    0.4 &              0.4 &              0.3 \\
    Scamander-G17 &  26.1 &          I &           + &    0.4 &              0.4 &              0.3 \\
        Slidr-I21 &   9.4 &         IV &           - &    0.5 &              0.4 &              0.3 \\
         Styx-G09 &  42.3 &          I &           + &    0.3 &              0.3 &              0.1 \\
         Svol-I21 &   7.8 &         II &           + &    0.3 &              0.3 &              0.4 \\
        \textbf{Sylgr-I21} &   8.7 &          I &           - &    0.3 &              0.5 &              0.2 \\
      Tri-Pis-B12 &  31.4 &        III &           - &    0.4 &              0.4 &              0.3 \\
    TucanaIII-S19 &  23.1 &        III &           - &    0.7 &              0.4 &              0.5 \\
       Turbio-S18 &  17.6 &        III &           - &    0.5 &              0.4 &              0.3 \\
  Turranburra-S19 &  32.9 &        III &           - &    0.7 &              0.3 &              0.3 \\
          Vid-I21 &  26.3 &        III &           - &    0.3 &              0.3 &              0.4 \\
    Wambelong-S18 &  19.5 &         IV &           + &    0.6 &              0.4 &              0.3 \\
  Willka\_Yaku-S18 &  34.4 &         IV &           - &    0.4 &              0.2 &              0.2 \\
         \textbf{Ylgr-I21} &  11.9 &          I &           - &    0.8 &              0.5 &              0.2 \\
    \end{longtable}}%
    \begin{tablenotes}
        \item \textbf{d}: distance of the stream from the MW. \textbf{Q}: Quadrant in the sky. \textbf{G}: Orbit's motion, prograde (+) or retrograde (-) with respect to the LMC-analog orbit in \mb{}. \textbf{e}: eccentricity of the stream orbit in \mb{}. Streams with encounter rate ratios of at least 2 between \mb{} and \mi{} are highlighted in bold. 
        \item $^*$Computed at $T = 0$ Gyr.  
        \item $^\dagger$ Missing proper motion information in {\fontfamily{qcr}\selectfont galstreams}. 
        \item Note that \cite{barry2023dark} demonstrated that the total subhalo population within a 50 kpc radius 5 Gyr ago was nearly twice as higher than the present-day population. Therefore our rates can be extrapolated to present day by appropriately scaling them down by a factor of 2. 
    \end{tablenotes}
\end{ThreePartTable}

\section{Discussion: the Milky Way in context} \label{sec:MW_cluster}


\begin{figure*}
    \includegraphics[width=\linewidth]{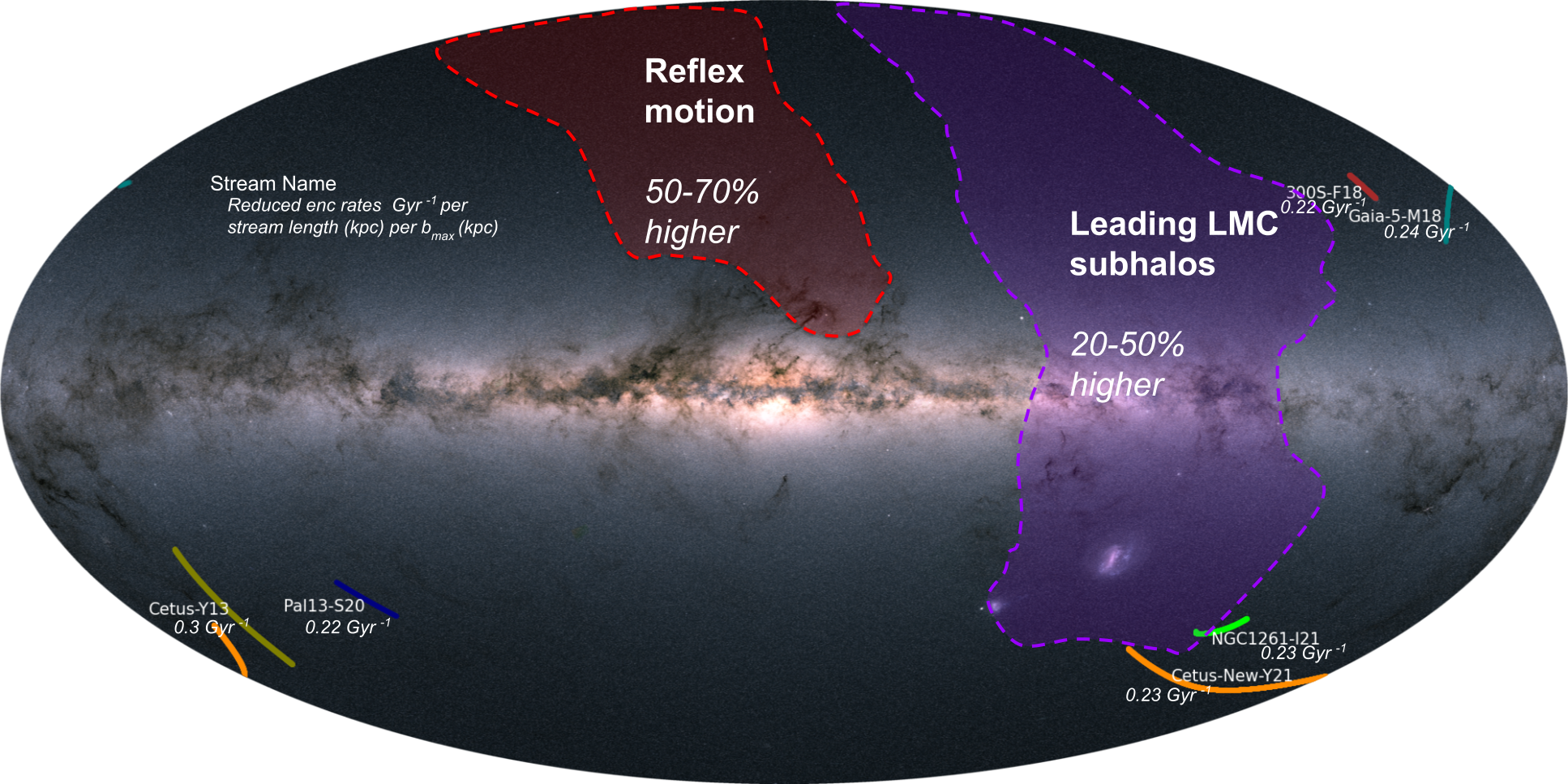}
    \caption{{\bf Regions with the highest encounter rates enhancements due to the LMC}, superimposed on the stellar flux map from Gaia DR2 \citep{gaia2018dr2}. Shaded regions on the map indicate higher overall probabilities of encounter rates, which have been determined using the top 1\% synthetic streams in \mb{}, rotated to align with the MW-LMC orientation at present-day. To aid visualization, the synthetic streams have been color-coded using a k-means clustering algorithm \citep{macqueen1967classification}, and shaded regions have been manually drawn. Also marked, the MW streams with the highest encounter rate enhancements located at distances greater than 20 kpc along with their reduced (by a factor of 2) rates per 10 kpc stream length and impact parameter in \mb{}. Streams within the shaded regions labeled as reflex motion (red) and subhalos brought in by the LMC (purple) are anticipated to have a higher probability of subhalo-stream interactions by as much as 50-70\% and 20-50\% respectively compared to the spherically averaged rates without accounting for the LMC contributed subhalos.}
    \label{fig:MW_cluster}
\end{figure*}

Currently, there are $\approx$100 observed streams in the MW. However, in the next decade, we expect to discover more streams, especially in the outer halo thanks to the upcoming surveys such as the LSST \citep{ivezic2019lsst}. Furthermore, we will have a multidimensional view of the phase-space and kinematics of the streams with unprecedented accuracy. As such, the search for disturbances from DM subhalos can be done systematically across large areas of the sky. With the insight from previous sections, we can anticipate specific areas in the MW sky where the likelihood of finding streams perturbed by subhalos is higher. 

In Fig.~\ref{fig:MW_cluster}, we highlight the specific regions in the sky with an increased probability of detecting stream--subhalo interactions. These regions are identified by analyzing the sky positions of synthetic streams at $T = 0$ Gyr in \mb{}, rotating them such that the LMC-analog matches the real LMC location in the MW, considering the expected changes in the rates with and without the contributed subhalos. These regions are shaded manually by running a k-means clustering algorithm on the stream locations \citep{macqueen1967classification}, dividing the stream positions in the Galactocentric coordinates into three distinct groups: one group for the background, another for the reflex motion, and a third for the LMC subhalos and DM wake categories, while also removing any outliers.   

The red-shaded region is the polar opposite region to the LMC, where the dominant effect on stream--subhalo interactions is the reflex motion response, both in the position and velocity. The purple-shaded region indicates a higher subhalo density due to the presence of LMC subhalos in the leading debris and the LMC itself. Streams located within these shaded regions are expected to have a higher likelihood of encounters with subhalos. Our estimates based on  Sec.~\ref{sec:results} suggest an enhancement ranging from 50\% to 70\% (depending on the stream's orbital orientation (Fig.\ref{fig:enc_rat_reg})) in the reflex motion region (red shaded area) and from 20\% to 50\% in the region influenced by the leading LMC subhalos (see Fig.\ref{fig:enc_rat_m12b_mu_change},\ref{fig:enc_rat_diff}) (purple shaded area) with respect to the spherically averaged rates without accounting for the LMC contributed subhalos. 

Additionally, we include the MW streams orbiting beyond 20~kpc from the galactic center, with encounter rates greater than the $90^\textrm{th}$ percentile limit (4.4 encounters per Gyr) based on our calculations in \mb{} (see Table~\ref{tab:enc_rate_streams}). We also list their corrected for present-day (reduced by a factor of 2) encounter rates per stream length of 10~kpc and impact parameter. 

We predict that streams such as NGC1261-I21 \citep{ibata2021charting}, Scamander-G17 \citep{grillmair2017tails}, Sangarius-G17 \citep{grillmair2017tails}, AAU-AliqaUma-L21 \citep{li2021broken}, Cetus-Y13 \citep{yam2013cetus}, and Gaia-5-M18 \citep{malhan2018streamfinder} have the highest likelihood of subhalo encounters in the outer regions of the stellar halo.

These predictions can be valuable for observing strategies aimed at detecting spurs, kinks, and gaps in streams during future surveys, particularly in the outer regions of the halo. Surveys such as the Vera Rubin Observatory \citep{ivezic2019lsst}, Nancy Grace Roman Space Telescope \citep{spergel2013wide, spergel2015wide}, DESI \citep{desicollaboration2016desi}, WEAVE \citep{dalton2012weave}, 4MOST \citep{de20194most}, Subaru PFS \citep{takada2014extragalactic}, among others, can prioritize regions in the sky where streams are more likely to have encounters with subhalos. It's important to note that caution is required when interpreting these values as calibration benchmarks for specific stream regions. Instead, they offer insight into the potential systematic uncertainties that can arise from departures from equilibrium assumptions in these measurements.    

\section{Conclusions} \label{sec:conc}

In this paper, we have investigated the impact of a massive satellite on the encounter rates between stellar streams and dark matter subhalos. Leveraging two representative systems from the FIRE-2 hydrodynamical zoom-in simulations—--one featuring an LMC-analog and the other lacking any massive satellite--—we address two key questions. First, we evaluated how the LMC-analog's subhalo population contributes to enhancing encounter rates with host stellar streams building on the work of \cite{barry2023dark} using the same simulations. Second, we investigated how the host halo's response influences encounter rates galaxy-wide. 

Our findings indicate a general decrease in encounter rates with increasing pericentric distance, independent of the eccentricity of stream's orbit. The presence of a massive satellite, such as the LMC-analog, introduces an {anisotropic boost}, resulting in varying encounter rates for streams across different regions of the sky relative to the satellite. Key factors influencing these asymmetric effects are ranked as follows:

\begin{enumerate}
    \item \textbf{Mean Radial Motion:} The mean radial motion of subhalos with respect to streams, represented by  $\tilde\mu$, consistently enhances encounter rates by at least 30\% even in systems without the LMC-analog. Under the influence of an LMC-analog, {localized boosts can jump up} to 70\% as observed in Q.I, compared to the spherically symmetric and $\tilde\mu=0$ rates (Fig.~\ref{fig:enc_rat_m12b_mu_change}). 
    
    \item \textbf{Contribution from the LMC-analog subhalos}: The LMC-analog brings its own set of subhalos with unique phase-space orbits, as depicted in Fig. \ref{fig:sat_char_m12b}. {These subhalos can boost the encounter rate up to 10--40\% in Q.III and Q.IV, aligned with the LMC-analog's motion (Fig.~\ref{fig:enc_rat_diff}, ~\ref{fig:enc_rat_m12c}).} This boost may be much higher in the MW today (appendix~\ref{app:m12c_enc}). \cite{barry2023dark} showed that the number density can be boosted up to a factor of 2 due to the LMC--mass satellites.
    
    \item \textbf{Orbital Alignment:} Streams in retrograde orbits with respect to the LMC-analog, {particularly} located in the outer halo and opposite hemisphere {from the analog} exhibit enhanced rates up to 30\%, with the most substantial boosts occurring in Q.I compared to prograde streams (Fig.~\ref{fig:enc_rat_reg}). 
\end{enumerate}

In conclusion, our investigations reveal that streams situated in proximity to the LMC and in regions opposite to it are more likely to interact with subhalos, potentially leading detectable morphological changes. Building upon this insight, we have identified regions within the MW--LMC celestial sphere likely to display signatures of stream--subhalo interactions, as depicted in Fig.~\ref{fig:MW_cluster}. Detecting such signatures and constraining the perturbing subhalo masses will offer valuable insights into allowed DM models with specific subhalo mass scales. Furthermore, our results emphasize the complexities introduced by the presence of a massive satellite such as the LMC, suggesting that the MW might not be an ideal laboratory for constraining the subhalo mass function due to the need to consider departures from the equilibrium assumptions. In this context, finding extra-galactic streams orbiting in galaxies in a state of dynamical equilibrium present promising opportunities for precise constraints on the DM subhalo mass function. With upcoming missions like the Roman Space Telescope \citep{spergel2013wide, spergel2015wide}, we anticipate the detection of such streams in other galaxies \citep{pearson2022hough, aganze2023prospects}, enhancing our ability to derive valuable insights into the nature of dark matter.

\begin{acknowledgments}
The authors thank {the anonymous referee}, Denis Erkal, Gurtina Besla, Kathryn V. Johnston, and Adrian Price-Whelan for valuable discussions that shaped this paper. 

AA and RES acknowledge support from the Research Corporation through the Scialog Fellows program on Time Domain Astronomy, from NSF grant AST-2007232, and from NASA grant 19-ATP19-0068. RES is supported in part by a Sloan Fellowship. ECC acknowledges support from NASA through the NASA Hubble Fellowship Program grant HST-HF2-51502 awarded by the Space Telescope Science Institute, which is operated by the Association of Universities for Research in Astronomy, Inc., for NASA, under contract NAS5-26555. AW and MB received support from: NSF via CAREER award AST-2045928 and grant AST-2107772; NASA ATP grant 80NSSC20K0513; HST grants AR-15809, GO-15902, GO-16273 from STScI.   

This research is part of the Frontera computing project at the Texas Advanced Computing Center (TACC). Frontera is made possible by the National Science Foundation award OAC-1818253. Simulations in this project were run using Early Science Allocation 1923870, and analyzed using computing resources supported by the Scientific Computing Core at the Flatiron Institute. This work used additional computational resources of the University of Texas at Austin and TACC, the NASA Advanced Supercomputing (NAS) Division and the NASA Center for Climate Simulation (NCCS), and the Extreme Science and Engineering Discovery Environment (XSEDE), which is supported by National Science Foundation grant number OCI-1053575.

AA would like to express sincere gratitude to ChatGPT for its invaluable language assistance (and for writing its own acknowledgment), and to OpenAI for providing the platform and technology. 

\end{acknowledgments}

\software{Astropy (\citealt{astropy:2013}, \citeyear{astropy:2018}, \citeyear{astropy:2022}), IPython \citep{ipython}, Matplotlib \citep{matplotlib}, Numpy \citep{numpy}, Pandas \citep{pandas2}, Scipy \citep{scipy}, Healpy \citep{healpy2019}, \texttt{consistent-trees} \citep{behroozi2012consistent}, 
\texttt{rockstar} \citep{behroozi2012rockstar}, \texttt{halo\_analysis} \citep{2020ascl.soft02014W},
\texttt{gizmo\_analysis} \citep{2020ascl.soft02015W}, \texttt{Galstreams} \citep{mateu2023galstreams}, {\fontfamily{qcr}\selectfont AGAMA} \citep{vasiliev2019agama}. CMasher \citep{cmasher}} 

\bibliography{ref}
\bibliographystyle{aasjournal}

\appendix

\section{Rotational transformation of the LMC-analog location to the real LMC} \label{app:lmc_rot}

{The LMC-analog in \mb{} is located at $\vec{x}_\textrm{analog} = (-23.3, -26.6,  13.7)$~kpc at $T_\textrm{peri}$ in the principal axes frame. However, we establish as the ``rotated axes'', the ones that align the position unit vector of the LMC-analog in \mb{} at its first pericentric passage with the position unit vector of the \emph{real} LMC in the MW at its first pericentric passage, located at $\vec{x}_\textrm{LMC} = (2.3, -20.2, -41.1)$~kpc, based on the orbits presented in \citet{garavito2019hunting}.   

We compute the rotation matrix $R_\textrm{LMC}$, using Rodrigues' rotation formula such that $\hat{x}_\textrm{analog} = R_\textrm{LMC}^T\hat{x}_\textrm{LMC}$ and:
\begin{equation}
    R_\textrm{LMC} = I + [v]_\times + \frac{{[v]^2}_\times}{1 + \hat{x}_\textrm{LMC}\cdot\hat{x}_\textrm{analog}}
\end{equation}
where $I$ is the identity, $v$ is given by $\hat{x}_\textrm{LMC} \times \hat{x}_\textrm{analog}$, and  $[v]_\times$ is the skew symmetric cross-product of $v$. 

For the specific values of $\vec{x}_\textrm{LMC} = (2.3, -20.2, -41.1)$~kpc and the analog in \mb{} $\vec{x}_\textrm{analog} = (-23.3, -26.6,  13.7)$~kpc, the rotation matrix is given by:

\begin{equation}
R_\textrm{LMC} = \begin{bmatrix}
  0.60508764 & -0.1328807 & 0.78499151\\
  -0.74603358 & 0.24968852 & 0.6173245\\
  -0.27803387 & -0.95916545 & 0.05194997\\
\end{bmatrix}  
\end{equation}

\section{Gaussianity Analysis of Subhalos' Radial Velocity Distributions}\label{app:gaussianity}

\begin{figure}[ht]
    \centering
    \includegraphics[width=0.5\linewidth]{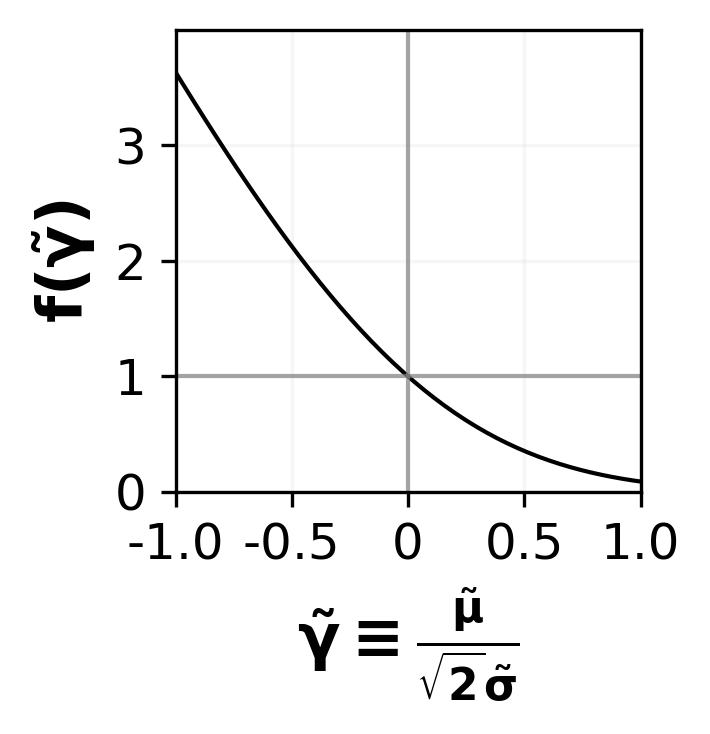}
    \caption{The anisotropic boost factor $f(\tilde\gamma)$ in eq.~\ref{eq:num_gauss} that is dependent on the Gaussian parameters of the cylindrical radial velocity distribution of subhalos with respect to a stream as a function of a typical $\tilde\gamma$, i.e is function of mean radial velocity ($\tilde\mu$) and velocity dispersion ($\tilde\sigma$) of subhalos in the stream-centric coordinates. Theoretically the factor converges to 0 with $\tilde\gamma \gg 0$ and diverges to infinity for $\tilde\gamma \ll 0$. $f(\tilde\gamma)$ strongly depends upon $\tilde\mu$ and $\tilde\sigma$ and hence assuming $\tilde\mu = 0$ can bias the estimates of encounter rates.}
    \label{fig:gamma_scal_fact}
\end{figure}

In this appendix, we investigate the Gaussianity assumption underlying the Probability Density Function (PDF) for the subhalos' radial velocity $P(\widetilde{v}_r$) in stream-centric coordinates. This assumption plays a crucial role in deriving an analytical model for the subhalo-stream encounter rates, as expressed in eq.~\ref{eq:num_gauss}.   

Fig.~\ref{fig:gamma_scal_fact} shows the {anisotropic boost factor}, $f(\tilde\gamma)$ from eq.~\ref{eq:num_gauss}, as a function of $\tilde\gamma$. The boost factor approaches 0 for a large positive mean velocity (i.e., all the subhalos moving away from the stream) and diverges to infinity for a large negative mean velocity (all the subhalos approaching the stream in the finite time interval). In the MW, the velocity dispersion of the outer halo is of order  $\sigma = 100$ km s$^{-1}$ \citep{deason2012cold, cohen2017outer}  and the expected mean velocity of $\tilde\mu = 35$ km s$^{-1}$ \citep{erkal2019total, petersen2020reflex} . For a MW stream in a circular orbit the anisotropic boost factor will scale the encounter rates to 1.75 or 0.5 depending on whether $\tilde\mu$ is locally negative or positive in the stream's frame. Notably, a negative $\tilde\mu$ value leads to a substantial rate enhancement than a positive $\tilde\mu$ leads to a reduction. For instance, a $\tilde\gamma$ of -0.25 results in a 50\% boost in encounter rates, while a +0.25 value leads to a 35\% decrease. For halos with no massive satellites, \cite{barry2023dark} showed that in the galactocentric frame at present day, the orbital velocity vectors of subhalos are generally isotropic at all distances.


\begin{figure}
    \includegraphics[width=\textwidth,height=\textheight,keepaspectratio]{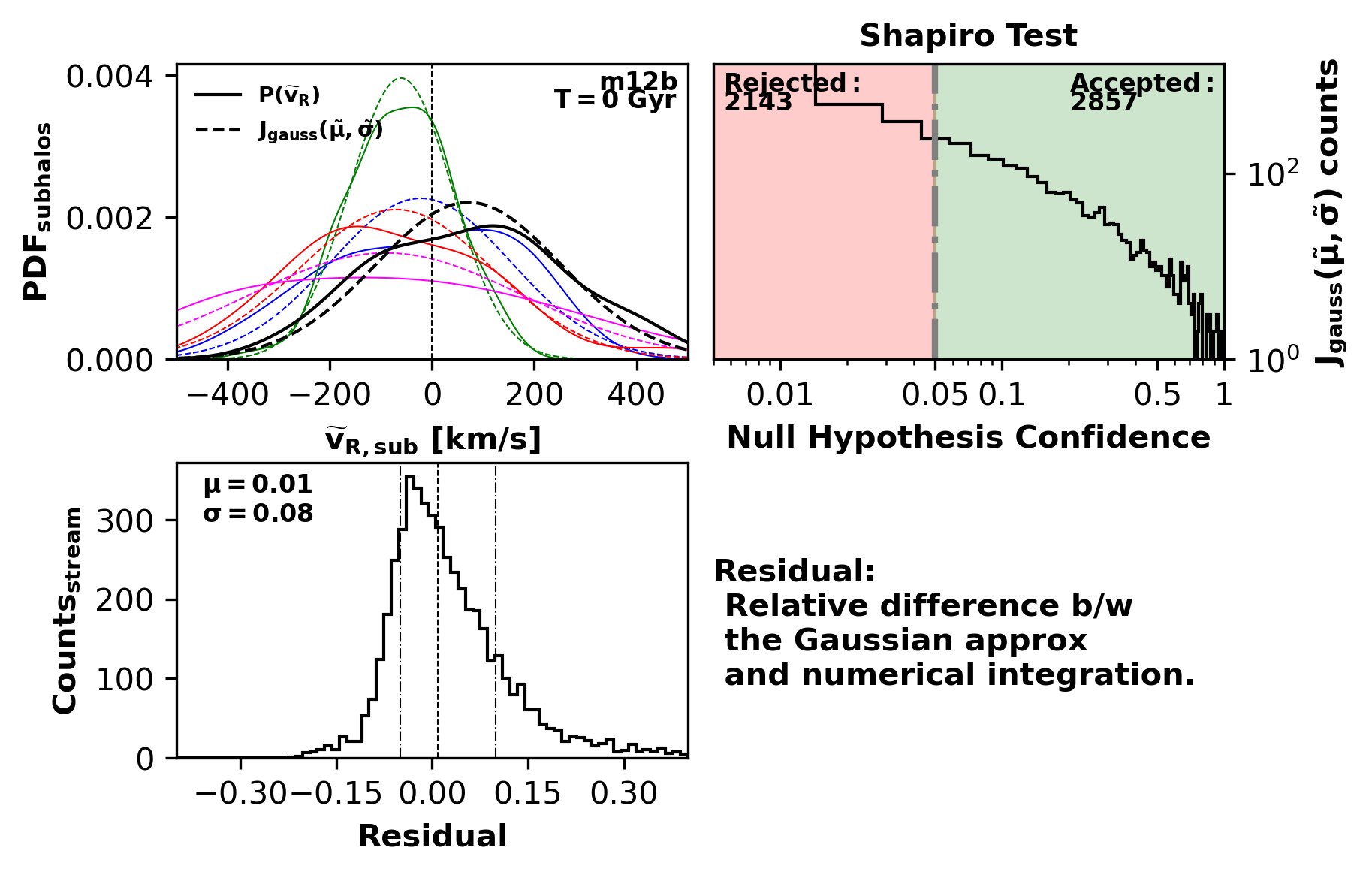}
    \caption{Left: Probability density function (PDF) of the cylindrical radial velocity $P(\widetilde{v}_R)$ in stream-centric coordinates (solid lines), along with corresponding Gaussian distribution fits $J_\textrm{gauss}(\tilde{\mu},\tilde{\sigma})$ (dashed lines) for a representative sample of MW stream orbits injected in \mb{}, illustrated with different colors at $T = 0$ Gyr. This comparison visually attests to the adequacy of Gaussian approximation.
    Right: Histogram of computed p-values obtained through the Shapiro-Wilk test \citep{shapiro1965analysis}, which assesses the null hypothesis of normal distribution. Applied to the PDF of subhalos' cylindrical radial velocity $P(\widetilde{v}_R)$ in stream-centric coordinates for the 5000 synthetic stream orbits in \mb{} at $T = 0$ Gyr. The red-shaded region indicates the 0.05 significance threshold, below which the null hypothesis is rejected. Conversely, green-shaded regions denote acceptance. Over 50 \% of PDFs analyzed follow Gaussianity from the accepted number of samples. {Bottom: Distribution of residuals calculated between the Gaussian approximation of eq.~\ref{eq:num_gauss} and the exact numerical integration (eq.~\ref{eq:I_vR}) without any assumptions on the cylindrical radial velocities of subhalos' distribution in stream-centric coordinates. Residuals are computed at $T=0$ Gyr for the 5000 synthetic streams in \mb{}. The dashed line represents the 50th quantile, and dashed-dotted lines mark the 16th and 84th quantiles. Most errors in assuming the approximation are within 10\%.}\label{fig:gauss_str}}
\end{figure}

We examine the validity of assuming a Gaussian distribution $J_\textrm{gauss} (\tilde\mu,\tilde\sigma)$ for $P(\widetilde{v}_R)$. It is noteworthy that the PDFs subjected to analysis in this section exclusively concern subhalos positioned around streams, in accordance with the description provided in Sec.~\ref{sec:anly_setup} within the context of the \mb{} at $T = 0$ Gyr. This specific choice is motivated by the fact that the LMC-analog's first pericentric passage induces significant perturbations on the Gaussianity assumption, owing to its pronounced dipolar moment \citep{petersen2020reflex, cunningham2020quantifying}.

The left panel of Fig.~\ref{fig:gauss_str} shows $P(\widetilde{v}_R)$ (solid lines) and their corresponding Gaussian fits (dashed lines) for a representative set of MW streams (varied colors). Notably, the means of different PDFs and their fits display similar trends. It's important to note that this mean value can shift, either towards more negative or positive values, thereby leading to the conditions $f(\gamma) \geq 1$ or $f(\gamma) \leq 1$ respectively. This shift's potential influence on the scaling factor behavior is highlighted. The alignment between Gaussian distributions and actual PDFs visually confirms the Gaussian fitting assumption.

For a quantitative analysis, we perform numerical integration for the 5000 synthetic streams at $T=0$ Gyr in \mb{}, assuming a Gaussian approximation (see equation~\ref{eq:num_gauss}), hereafter labeled as $\mathcal{I}_\textrm{gauss}$, and conducting the exact numerical integral ($\mathcal{I}_\textrm{num}(\tilde{v}_{R})$) without any assumption (see equation~\ref{eq:dNenc_raw} in Sec.~\ref{sec:anly_setup}). We compute the residual as $\mathcal{I}_\textrm{gauss} / \mathcal{I}_\textrm{num} - 1$ to assess the errors introduced by the approximation.

The bottom panel of Fig.~\ref{fig:gauss_str} shows the resulting residual distribution for the 5000 synthetic streams in m12b at $T = 0$ Gyr. The dashed line represents the 50th quantile, and the dashed-dotted lines mark the 16th and 84th quantiles. The majority of errors fall within 10\% of the expected values, with the mean and standard deviation of the distribution being 0.01 and 0.08, respectively. This translates to an average error of 1\% with a deviation of 8\%. We also note that the majority of higher errors ($\geq 10 \%$) occur for streams close to the galactic center and the disk stemming from the low subhalo population. These results highlight the accuracy of the Gaussian approximation method, specially compared to other assumptions used in such analysis.

For a qualitative evaluation, we test the Gaussian assumption using the Shapiro-Wilk test \citep{shapiro1965analysis} on the PDFs of 5000 synthetic streams. This statistical test evaluates the null hypothesis of Gaussian distribution for each $P(\widetilde{v}_r)$. 

The right panel of Fig.~\ref{fig:gauss_str} depicts a histogram of computed p-values (Null hypothesis confidence) from the Shapiro-Wilk test. The red-shaded region identifies the 0.05 significance threshold, marking p-values below which the null hypothesis is rejected. Among the analyzed distributions, 2143 exhibit p-values below 0.05 (non-Gaussian behavior), while 2857 distributions show p-values above 0.05 (green region), reinforcing the Gaussian assumption as over 50\% of the distributions have p-values in the green region. 

Remarkably, the p-value counts demonstrate a logarithmic decrease, highlighting a gradual transition from non-Gaussian to Gaussian behavior in the PDFs. This observation further endorses the Gaussian approximation's appropriateness for the PDFs. While the numerical 1-D integrals are efficient, the assumption of Gaussianity remains valid and imposes no significant constraints compared to other approximations commonly employed in such analyses.

\section{Encounter rates in another LMC-analog (m12c)}\label{app:m12c_enc}


\begin{figure}[ht]
    \includegraphics[width=\textwidth,height=\textheight,keepaspectratio]{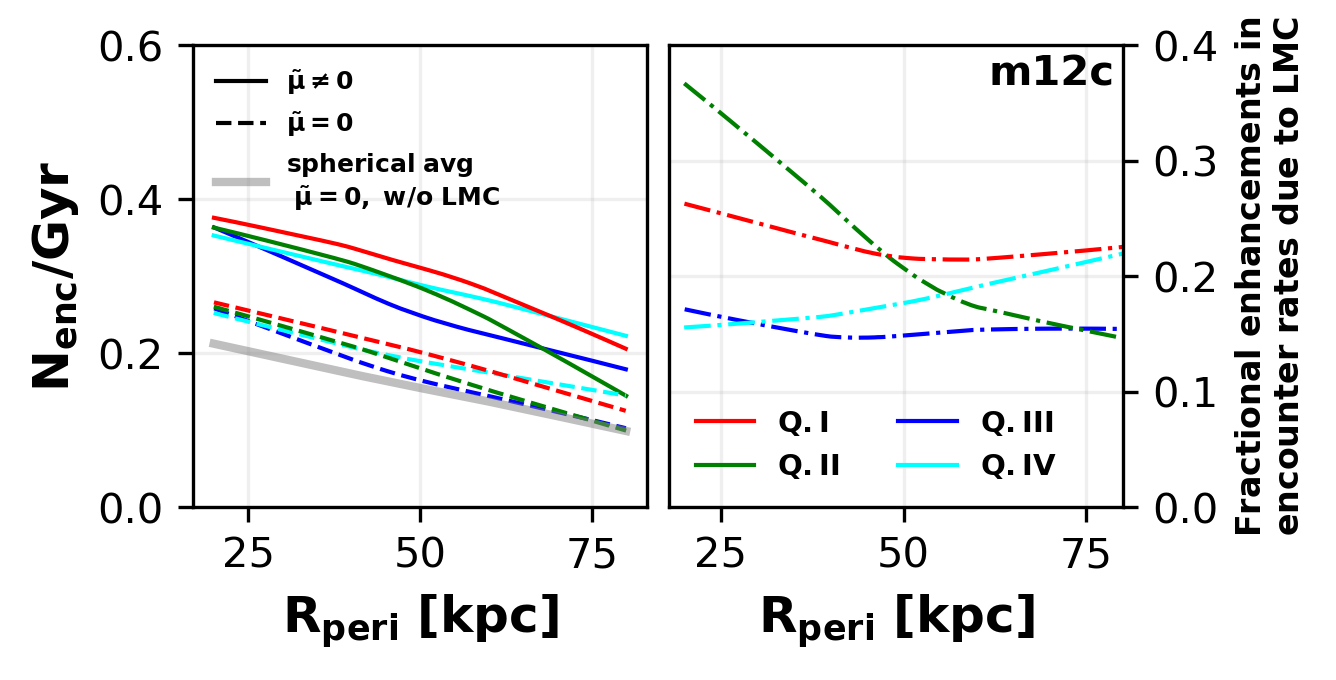}
    \caption{Encounter rates in Sec.\ref{sec:enc_rat_synthetic} for m12c, assuming $T = 0$ Gyr when the LMC-analog is at a distance of 50 kpc from the host center. Left: Estimated number of encounters per Gyr per unit stream length $\ell_s$ and maximum impact parameter $\mathrm{b_{max}}$ in m12c with $\tilde\mu$ (solid) and w/o $\tilde\mu$ (dashed) plotted as a function of the stream's pericenter distance at T=0 Gyr and four quadrants (marked by colors) (compared to Fig.~\ref{fig:enc_rat_m12b_mu_change}). The gray line represents the range of encounter rates for m12c with $\tilde\mu = 0$ and without contributed subhalos, averaged over all quadrants. The highest encounter rates occur in Q.I and Q.IV (where the LMC-analog orbits in the integration time frame). Right: Fractional enhancements in the encounter rates due to contributed subhalos from the LMC-analog for m12c (compared to Fig.~\ref{fig:enc_rat_diff}). Q.I and Q.IV exhibit a consistent 20\% increase in the encounter rates in the outer halo due to a higher subhalo density around the analog. In contrast, Q.II shows the highest enhancements in the inner regions (up to 35\%) due to leading arm of the contributed subhalos.}\label{fig:enc_rat_m12c}
\end{figure}

In this appendix, we repeat our analysis from Sec.~\ref{sec:enc_rat_synthetic}. We achieve this by injecting and integrating 5000 synthetic stream orbits, as described in Sec.~\ref{Sec:injected_st} in the context of m12c, which has \emph{another} LMC-analog, with the first pericentric passage at about 1 Gyr before the present day, at a much closer distance of 18 kpc from the host center. Following \citet{barry2023dark}, we assume $T = 0$ Gyr (first pericentric passage), when the LMC-analog is 50 kpc away from the center. The LMC-analog's orbit largely lies in the disc plane until the $T = 0.2$ Gyr. During this time, the analog starts in Q.IV and moves inwards with the first pericenter ($T = 0$ Gyr) into Q.I. The actual pericenter occurs in Q.II at $T = 0.15$ Gyr. We integrate our synthetic stream orbits from -0.4 Gyr to 0 Gyr (in contrast to -0.4 Gyr to 0.1 Gyr in the \mb{} case) and do not anticipate significant north--south asymmetry (collective response) or the emergence of reflex motion during this period, given that the LMC remains in the galactic disc plane.

The left panel in Fig.~\ref{fig:enc_rat_m12c} shows the average encounter rates of synthetic streams as a function of their pericentric distances in different quadrants (color-coded) within the context of m12c. A comparison with the encounter rates in \mb{}(Fig.~\ref{fig:enc_rat_m12b_mu_change}) reveals an overall reduction by a factor of 1.6. This decrease is due to a higher number of subhalos in the MW 5 Gyr before the present day compared to 1 Gyr ago \citep[see Fig. 4 in][]{barry2023dark}. Notably, Q.I and Q.IV exhibit the highest encounter rates, aligning with the LMC-analog's orbit and the expected DM wake. Q.II also displays some increased rates in the inner regions.

The right panel in Fig.~\ref{fig:enc_rat_m12c} plots the fractional enhancements in $\mathrm{N_{enc}/Gyr}$ due to the contributed subhalos in m12c. Even though both the LMC-analogs (in m12c and \mb{}) start with the same number of subhalos, the overall enhancements in m12c are higher than \mb{} (Fig.~\ref{fig:enc_rat_diff}) due to decreased MW subhalo population at later times. Q.I and Q.II exhibit a consistent enhancement of 30\% or more in the inner regions due to the leading arm of the contributed subhalos tracing the analog's future orbit, while Q.I and Q.IV both experience over 20\% increase in the outer halo. The total fractional enhancement for the real LMC is more accurate for m12c, given its LMC-analog's proximity to the present day, it's important to note that the quadrant and distance dependencies are highly sensitive to the orbit and pericentric distance, making \mb{} more suitable for predicting regions of enhancements from the real LMC.
\end{document}